\RequirePackage{fix-cm}
\documentclass[twocolumn,epjc3]{svjour3}  

\usepackage{amsmath}
\usepackage{lipsum} %
\smartqed  
\usepackage{orcidlink}
\RequirePackage{graphicx}
\usepackage{graphicx} %
\usepackage{subcaption} %
\journalname{Eur. Phys. J. C}
\begin{document}

\title{Holographic CFT phase transitions and criticality for
charged Gauss-Bonnet AdS black holes in the ensemble at fixed $(C, \mathcal{V}, \tilde{Q}, \tilde{\mathcal{A}})$
}

\author{Limin Zeng\orcidlink{0009-0003-5703-2687}\thanksref{e1,addr1,addr2,addr3}
}

\thankstext{e1}{e-mail: zenglimin25@mails.ucas.ac.cn}


\institute{School of Fundamental Physics and Mathematical Sciences,\\Hangzhou Institute for Advanced
Study, UCAS, Hangzhou 310024, China \label{addr1}
           \and 
           Institute of Theoretical Physics, Chinese Academy of Sciences, Beijing 100190, China \label{addr2}
           \and
           University of Chinese Academy of Sciences, Beijing 100049, China
           \label{addr3}
}

\date{Received: date / Accepted: date}

\maketitle

\begin{abstract}
We study the holographic dual of the extended thermodynamics of spherically symmetric, charged Gauss-Bonnet AdS black holes in the context of the AdS/CFT correspondence. Compared to Einstein's theory of gravity, Gauss-Bonnet gravity introduces higher-order curvature terms. The coupling constants of these higher-order curvature terms $\alpha$ can serve as new thermodynamic quantities, which will also be dual to thermodynamic quantities on the boundary CFT, a feature not present in the CFT dual to Einstein's gravity previously. Based on the holographic dictionary, we studied the critical behavior and phase transition of the CFT description of the charged Gauss-Bonnet black holes in $d=4$ and $d=5$ in the ensemble at fixed $(C, \mathcal{V}, \tilde{Q}, \tilde{\mathcal{A}})$. The interesting behaviour of free energy stems from the fact that the constraints we introduced to handle the gravitational constant on CFT and the AdS radius differ from conventional approaches. Using the criticality equation, we numerically found the critical points of the zeroth-order and first-order phase transition for $\tilde{\mathcal{A}}$. The relationships between conjugate thermodynamic pairs (equation of state) were also examined. In the case of the $p-\mathcal{V}$, $\tilde{T}-\tilde{S}$ and $\tilde{\Phi}-\tilde{Q}$ conjugate pairs, characteristics that are analogous to the first-order phase transition of van der Waals fluids were found.   
\end{abstract}

\section{Introduction}
\label{intro}
Analogous to the hydrogen atom's pivotal role in the development of quantum mechanics during the last century, black holes have emerged as fundamental testing grounds in contemporary quantum gravity research. With the introduction of a negative cosmological constant modeled as an ideal fluid, extended black hole thermodynamics (often termed black hole chemistry) has been extensively investigated over recent decades.

Within Einstein's theory of gravity, two quintessential quantum manifestations emerge: (i) the black hole entropy demonstrates direct proportionality to the event horizon area ($A/4$ law)\cite{Bekenstein:1973ur}, and (ii) the Hawking radiation temperature scales with the surface gravity at the horizon\cite{HW1975}. See \eqref{eq:1.1}.
\begin{equation}
S = \frac{A}{4G_N},\quad T = \frac{\kappa}{2\pi}
\tag{1.1} \label{eq:1.1}
\end{equation}
Notably, black holes in asymptotically AdS spacetimes exhibit rich phase transition phenomena, including first-order phase transitions between thermal AdS and black hole\cite{Hawking:1982dh} corresponding to confinement/deconfinement of the dual quark
gluon plasma\cite{Witten:1998zw}, as well as critical behavior resembling van der Waals fluid liquid-gas phase transitions in charged AdS black holes\cite{Kubiznak:2012wp}, and so on.
In the more general framework of LoveLock gravity, coupling parameters of higher-curvature terms introduce systematic modifications to black hole thermodynamic quantities~\cite{lovelockbulk}. The leading-order correction effects are particularly well-characterized through Gauss-Bonnet gravity~\cite{Wei:2014hba}.

Furthermore, the gauge/gravity duality (AdS/CFT correspondence), since its seminal proposal by Maldacena \cite{Maldacena:1997re}, has become one of the most actively studied approaches in quantum gravity. The combination of black hole chemistry with holographic principles has recently given rise to the nascent field of "holographic black hole chemistry" (see ref. ~\cite{Mann:2025xrb} for a comprehensive review). This discipline focuses on establishing precise thermodynamic correspondences between bulk black holes and their dual conformal field theories (CFTs) at the boundary. 

The criticality of the CFT dual to the charged AdS black hole in Einstein gravity has been thoroughly investigated \cite{Cong:2021jgb}. However, upon incorporating corrections from Gauss-Bonnet gravity, higher-order curvature coupling coefficient $\alpha$ and their conjugate thermodynamic quantity $\mathcal{A}$ are introduced into the original conjugate thermodynamic framework of the dual CFT\cite{Wei:2014hba}. Ref.~\cite{Sadeghi:2024ish} demonstrates that for charged AdS black holes in Gauss-Bonnet gravity, the extended first law of thermodynamics in the dual CFT holds only in dimensions $d=4$ and $d=5$. And \cite{Sadeghi:2024ish} and \cite{Qu:2022nrt} study the holographic CFT phase transitions and criticality for charged Gauss-Bonnet AdS black
holes in the ensemble at fixed $(C, \mathcal{V}, \tilde{Q}, \tilde{\alpha})$. Besides, ref.~\cite{Yang:2024krx} provides a detailed study of the thermodynamic properties of neutral Gauss-Bonnet AdS black holes. What's more, a recent study in ref. \cite{panigrahi2025} investigated the thermodynamics and phase structure of a deformed AdS-Schwarzschild black hole generated via the gravitational decoupling method as a minimal geometric deformation, and also demonstrated various exotic phase transition behaviors of the bulk spacetime and dual CFT, exhibiting certain similarities to our work.

The purpose of the present paper is to investigate the holographic dual of extended
thermodynamics of $d=4$ and $d=5$ Gauss-Bonnet charged AdS black holes in the ensemble at fixed $(C, \mathcal{V}, \tilde{Q}, \tilde{\mathcal{A}})$.
In Section~\ref{sec:2}, we review the thermodynamic quantities and relations of the CFT dual to Gauss-Bonnet gravity briefly. In Section~\ref{sec:3}, we will discuss the phase transition behavior and critical phenomena of the CFT in the ensemble at fixed $(C, \mathcal{V}, \tilde{Q}, \tilde{\mathcal{A}})$. These peculiar phenomenons of free energy stem from the artificial constraints that must be introduced when the system is forcibly fitted into the traditional free energy framework as we can see. Besides, using the criticality equation, we will numerically determine the critical point of the zeroth-order phase transition and the first-order phase transition for $\tilde{\mathcal{A}}$.
In section~\ref{sec:4}, based on our current understanding, we will discuss relationships of conjugate thermodynamic pairs (equation of states) in the ensemble at fixed $(C, \mathcal{V}, \tilde{Q}, \tilde{\mathcal{A}})$. $p-\mathcal{V}$, $\tilde{T}-\tilde{S}$ and $\tilde{\Phi}-\tilde{Q}$ conjugate pairs show characteristics similar to the first-order phase transition of a van der Waals fluid. In section~\ref{sec:5}, we will summarize this paper. 

\section{Holographic thermodynamics of Gauss-Bonnet gravity}
\label{sec:2}

The AdS/CFT correspondence establishes a fundamental relationship between the thermodynamics of AdS black holes and that of the dual CFT\cite{Witten:1998zw}. In this section, we revisit the holographic dictionary for the extended thermodynamics of charged AdS black holes in Gauss-Bonnet gravity, as detailed in ref. ~\cite{Sadeghi:2024ish}.
\subsection{$d = 5$}

We begin with the action of the $d$-dimensional Einstein-Maxwell theory ($d \ge 5$) of Gauss-Bonnet gravity, incorporating the negative cosmological constant and Gauss-Bonnet term, it reads~\cite{Sadeghi:2024ish,Qu:2022nrt,Cai:2001dz,Yerra:2022alz} 
\begin{equation}
\begin{split}
S &= \frac{1}{16\pi G} \int d^d x \sqrt{-g}  R - 2\Lambda \\ &+ \alpha_{GB} \left( R_{\mu\nu\rho\sigma} R^{\mu\nu\rho\sigma} + R^2 - 4R_{\mu\nu} R^{\mu\nu} \right) - 4\pi G F_{\mu\nu} F^{\mu\nu} 
\end{split}
\tag{2.1} \label{eq:2.1}
\end{equation}
where the $\Lambda$ is defined by
\begin{equation}
\Lambda = -\frac{(d-2)(d-1)}{2\ell^2}, \quad P = -\frac{\Lambda}{8\pi G}.
\tag{2.2} \label{eq:2.2}
\end{equation}
$\Lambda$ is interpreted as the positive bulk pressure $P$ of the system~\cite{Kastor:2009wy,Dolan:2011xt,Dolan:2010ha,Kubiznak:2014zwa}.
In this work, We focus on the spherical topology of the horizon, meaning that $k = 1$ in some literatures' notation. 
And we define $\alpha=(d-4)(d-3)\alpha_{GB}$, with the dimension of $(length)^{2}$. The first law and Smarr relation can be expressed as follows\cite{Sadeghi:2024ish,Qu:2022nrt,Cai:2001dz,Yerra:2022alz}
\begin{equation}
\delta M = T \delta S + \Phi \delta Q + V \delta P + \mathcal{A} \delta \alpha,
\tag{2.3} \label{eq:2.3}
\end{equation}
\begin{equation}
M = \frac{d-2}{d-3} TS + \Phi Q - \frac{2}{d-3} VP + \frac{2}{d-3} \mathcal{A} \alpha.
\tag{2.4} \label{eq:2.4}
\end{equation}
$\mathcal{A}$ is the conjugate of $\alpha$. 

In the AdS/CFT context, the central charge $C$ of the CFT dual to the AdS bulk is related to its AdS radius $\ell$ as follows:
\begin{equation}
C = \frac{k \ell^{d-2}}{16\pi G}.
\tag{2.5} \label{eq:2.5}
\end{equation}
$k$ is a constant depending on details of the particular holographic
system\cite{Karch:2015rpa}. Besides, $C$ is proportional
to $N$ to some power, for example $C \propto
 N^2 $ for $SU(N)$ gauge theories with conformal symmetry. For a given CFT, the central charge 
$C$ is fixed. In the dual gravitational system, varying the AdS radius $\ell$
necessitates a corresponding adjustment of the gravitational constant $G$. More generally, when $C$ changes, both $\ell$ and 
$G$ must vary simultaneously. This represents one of the most significant distinctions between the holographic chemistry framework and traditional extended black hole thermodynamics. By using the relation \eqref{eq:2.5}, one can introduce boundary central charge $C$ into the first law\cite{Visser:2021eqk}. In our case, the mixed thermodynamics relation is as follows\cite{Sadeghi:2024ish,Qu:2022nrt}:
\begin{equation}
\delta M = T \delta S + \Phi \delta Q + V_{bb} \delta P + \mu_{bb} \delta C + \mathcal{A} \delta \alpha.
\tag{2.6} \label{eq:2.6}
\end{equation}
$\mu_{bb}$ is regarded as the chemical potential corresponding to the color charge or the central charge\cite{Visser:2021eqk}. For the expressions of specific thermodynamic quantities in\eqref{eq:2.6} , you can refer to \cite{Sadeghi:2024ish,Qu:2022nrt}.

Anyway, using the holographic dictionary, we can derive the CFT thermodynamics with a chemical potential for the central charge. This is what we are truly interested in. In fact, as it is pointed out in \cite{Sadeghi:2024ish}, under different dimensions, the holographic dictionaries for the Gauss-Bonnet gravity are different. But there is something in common under different dimensions:
\begin{equation}
\begin{split}
E = \frac{M}{\omega}, \quad 
\tilde{S} &= S, \quad 
\tilde{T} = \frac{T}{\omega}, \\ \quad 
\tilde{\Phi} = \frac{\Phi \sqrt{G}}{\omega \ell},  \quad 
\tilde{Q} &= \frac{Q \ell}{\sqrt{G}},\quad \omega= \frac{R}{\ell}.
\end{split}
\tag{2.7} \label{eq:2.7}
\end{equation}
$R$ is the curvature radius of the boundary. So the metric of the CFT, which exhibits conformal scaling invariance, can be expressed as follows
\begin{equation}
ds^2 = \omega^2 \left( -dt^2 + \ell^2 d\Omega_{d-2}^2 \right).
\tag{2.8} \label{eq:2.8}
\end{equation}
In this case, the volume of the CFT is given by
\begin{equation}
\mathcal{V} = \omega_{d-2} R^{d-2}.
\tag{2.9} \label{eq:2.9}
\end{equation}
$\omega_{d-2}$ is the volume of $(d - 2)$-dimensional sphere. All in all, using the scale transformations of \eqref{eq:2.3}\eqref{eq:2.4}\eqref{eq:2.5}, we can get,
\begin{equation}
\begin{split}
\delta\left(\frac{M}{\omega}\right) &= \frac{T}{\omega}\delta\left(\frac{A}{4G}\right) \\ &+ \left(\frac{M}{\omega} - \frac{TS}{\omega} - \frac{Q\Phi}{\omega} - \frac{\mathcal{A}\alpha}{\omega}\right)\frac{\delta(k\ell^{d-2}/G)}{k\ell^{d-2}/G} \\
&-\frac{M}{\omega(d-2)}\frac{\delta(\omega_{d-2}R^{d-2})}{\omega_{d-2}R^{d-2}} + \frac{\Phi\sqrt{G}}{\omega\ell}\delta\left(\frac{Q\ell}{\sqrt{G}}\right) \\&+ \frac{\mathcal{A}}{\omega\ell}\left(\ell\delta\alpha + (d-4)\alpha\delta\ell\right).
\end{split}
\tag{2.10} \label{eq:2.10}
\end{equation}
According to the last part of relation \eqref{eq:2.10}, we can only have the first law of CFT when $d = 4$ or $d = 5$\cite{Sadeghi:2024ish}. This is an uncommon fact, because within the framework of LoveLock gravity, the generalized black hole thermodynamic relations in the bulk can be extended to higher orders and higher dimensions\cite{lovelockbulk}.

For $d=5$, we should take the following scale transformation
\begin{equation}
\begin{split}
\tilde{\mathcal{A}} = \frac{\mathcal{A}}{\omega \ell}, \quad 
\tilde{\alpha} = \ell \alpha
\end{split}
\tag{2.11} \label{eq:2.11}
\end{equation}
Then \eqref{eq:2.10} can be written as
\begin{equation}
\begin{split}
\delta\left(\frac{M}{\omega}\right) &= \frac{T}{\omega} \delta\left(\frac{A}{4G}\right)\\ &+ 
\left( \frac{M}{\omega} - \frac{TS}{\omega} - \frac{Q\Phi}{\omega} - \frac{\mathcal{A}\alpha}{\omega} \right) \frac{\delta(k\ell^3/G)}{k\ell^3/G} \\
&- \frac{M}{3\omega} \cdot \frac{\delta(\omega_{3}R^3)}{\omega_{3}R^3} + 
\frac{\Phi\sqrt{G}}{\omega\ell} \delta\left( \frac{Q\ell}{\sqrt{G}} \right) + 
\frac{\mathcal{A}}{\omega\ell} \delta(\ell\alpha),\\
\rightarrow \delta E &= \tilde{T} \delta \tilde{S} + \mu \delta C - p \delta \mathcal{V} + \tilde{\Phi} \delta \tilde{Q} + \tilde{\mathcal{A}} \delta \tilde{\alpha}.
\end{split}
\tag{2.12} \label{eq:2.12}
\end{equation}
What's more, from \eqref{eq:2.7} and \eqref{eq:2.12}, we have,
\begin{equation}
\mu = \frac{1}{C} \left( E - \tilde{T} \tilde{S} - \tilde{\Phi} \tilde{Q} - \tilde{\mathcal{A}} \tilde{\alpha} \right), \quad p = \frac{E}{3\mathcal{V}}, \quad \mathcal{V}=\omega_{3}R^{3}.
\tag{2.13} \label{eq:2.13}
\end{equation}
This equation gives us the holographic Smarr relation:
\begin{equation}
E = \tilde{T} S + \tilde{\Phi} \tilde{Q} + \tilde{\mathcal{A}} \tilde{\alpha} + \mu C.
\tag{2.14} \label{eq:2.14}
\end{equation}
A noteworthy study is ref \cite{mancilla2025}, which proposes deriving a generalized Euler equation from the effective field theory formulation of perfect fluids. This equation is independent of the AdS/CFT correspondence and can naturally recover the Smarr formula for AdS black holes, thereby situating the physical interpretation of the Smarr formula within the framework of well-established physics.

It should be emphasized that \eqref{eq:2.14} no longer contains the volume-pressure term from the bulk.
It is convenient for us to introduce
the dimensionless parameters for $d=5$:
\begin{equation}
x \equiv \frac{r_+}{\ell}, \quad y \equiv \frac{\mathcal{V}^{1/3}\tilde{\alpha}}{3C}.
\tag{2.15} \label{eq:2.15}
\end{equation}
In this case, the thermodynamic quantities can be written in terms of $x$ and $y$ as follows (for simplicity, we set
$\omega_{3}=1$, the area of $(d - 2)$-dimensional unit sphere $\Sigma =1$ and $k=1$):
\begin{equation}
E = \frac{ \left(  \tilde{Q}^2 + 768\pi^2 C^2 x^6 + 768\pi^2 C^2 x^4 + 768\pi^2 C^2 x^2 \tilde{\mathcal{A}}y \right)}{256\pi^2 C  \mathcal{V}^{1/3} x^2},
\tag{2.16} \label{eq:2.16}
\end{equation}
\begin{equation}
\tilde{S} = {4\pi C  (x^3 + 6x\tilde{\mathcal{A}}y)},
\tag{2.17} \label{eq:2.17}
\end{equation}
\begin{equation}
\tilde{T} =  \frac{1}{\mathcal{V}^{1/3}}  \left( \frac{ -\tilde{Q}^2 + 1536\pi^2 C^2 x^6 + 768\pi^2 C^2 x^4}{1536\pi^3 C^2 x^3 (x^2 + 2\tilde{\mathcal{A}}y)} \right),
\tag{2.18} \label{eq:2.18}
\end{equation}
\begin{equation}
\tilde{\Phi} = \frac{1}{\mathcal{V}^{1/3}}  \left( \frac{ \tilde{Q}}{128\pi^2 c x^2} \right),
\tag{2.19} \label{eq:2.19}
\end{equation}
\begin{equation}
p = -\left( \frac{\partial E}{\partial \mathcal{V}} \right)_{\tilde{S}, \tilde{Q}, C, \tilde{\mathcal{A}}} = \frac{E}{3\mathcal{V}},
\tag{2.20} \label{eq:2.20}
\end{equation}
\begin{equation}
\mu = \frac{ \left( \tilde{Q}^2 + 768\pi^2 C^2 x^6 + 768\pi^2 C^2 x^4 + 768\pi^2 C^2 x^2 \tilde{\mathcal{A}}y \right)}{256\pi^2 C^2  {\mathcal{V}}^{1/3} x^2},
\tag{2.21} \label{eq:2.21}
\end{equation}
\begin{equation}
\tilde{\alpha} = y\tilde{\mathcal{A}}\ell^3.
\tag{2.22} \label{eq:2.22}
\end{equation}

\subsection{$d=4$}

However, only considering $d \ge 5$ is insufficient for our discusion here, as was explicitly pointed out in ref. ~\cite{Sadeghi:2024ish}. The $d=4$ Gauss-Bonnet coupling corresponds to the topological effect. In the framework of Gauss-Bonnet gravity, static and spherically symmetric black hole solutions are well-established in spacetime of dimension 
$d \ge 5$. The Gauss-Bonnet term, however, becomes topologically trivial in four dimensions, leading to its absence from the field equations and precluding the existence of Gauss-Bonnet black holes in this context. A breakthrough was achieved by Glavan and Lin \cite{Glavan:2019inb}, who circumvented this limitation through a novel rescaling of the Gauss-Bonnet coupling parameter 
$\alpha \to \alpha/(d-4)$, then $\alpha \equiv\alpha_{GB}$ followed by the 
$d \to 4$ limit, thereby obtaining a non-trivial four-dimensional black hole solution. Subsequent work generalized this solution to incorporate charged configurations within an AdS spacetime~\cite{Sadeghi:2024ish}.

Following the same discussion as for the case of $d=5$, and by adopting the following special scaling:
\begin{equation}
\tilde{\mathcal{A}} = \frac{\mathcal{A}}{\omega}, \quad 
\tilde{\alpha} = \alpha,
\tag{2.23} \label{eq:2.23}
\end{equation}
\eqref{eq:2.10} can be written as
\begin{equation}
\begin{split}
\delta\left(\frac{M}{\omega}\right) &= \frac{T}{\omega} \delta\left(\frac{A}{4G}\right) \\ &+ 
\left( \frac{M}{\omega} - \frac{TS}{\omega} - \frac{Q\Phi}{\omega} - \frac{\mathcal{A}\alpha}{\omega} \right) \frac{\delta(k\ell^2/G)}{k\ell^2/G} \\
&- \frac{M}{2\omega} \cdot \frac{\delta(\omega_{2}R^2)}{\omega_{2}R^2} + 
\frac{\Phi\sqrt{G}}{\omega\ell} \delta\left( \frac{Q\ell}{\sqrt{G}} \right) + 
\frac{\mathcal{A}}{\omega} \delta\alpha, \\
\rightarrow \delta E &= \tilde{T} \delta \tilde{S} + \mu \delta C - p \delta \mathcal{V} + \tilde{\Phi} \delta \tilde{Q} + \tilde{\mathcal{A}} \delta \tilde{\alpha}.
\end{split}
\tag{2.24} \label{eq:2.24}
\end{equation}
What's more, from \eqref{eq:2.7} and \eqref{eq:2.24}, we have,
\begin{equation}
\mu = \frac{1}{C} \left( E - \tilde{T} \tilde{S} - \tilde{\Phi} \tilde{Q} - \tilde{\mathcal{A}} \tilde{\alpha} \right), \quad p = \frac{E}{2\mathcal{V}}, \quad \mathcal{V}=\omega_{2}R^{2},
\tag{2.25} \label{eq:2.25}
\end{equation}
This equation also gives us the holographic Smarr relation \eqref{eq:2.14}. it is convenient for us to introduce
the dimensionless parameters for $d=4$:
\begin{equation}
x \equiv \frac{r_+}{\ell}, \quad y \equiv \frac{\sqrt{\mathcal{V}}\tilde{\alpha}}{8\pi C}.
\tag{2.26} \label{eq:2.26}
\end{equation}
In this case, the thermodynamic quantities can be written in terms of $x$ and $y$ as follows (for simplicity, we set
$\omega_{2}=1$ and $k=1$):
\begin{equation}
\begin{split}
E &= \frac{1}{32\pi C x \sqrt{\mathcal{V}}} \Bigg(
 \tilde{Q}^2 + 256\pi^2 C^2 x^4 \\&+ 256\pi^2 C^2 x^2 + 256\pi^2 C^2 \tilde{\mathcal{A}}xy
\Bigg),
\end{split}
\tag{2.27} \label{eq:2.27}
\end{equation}
\begin{equation}
\tilde{S} = {16\pi^2 C} \bigg( x^2 + 2\tilde{\mathcal{A}}xy \ln \bigg[ \frac{x}{{\tilde{\mathcal{A}}y}} \bigg] \bigg),
\tag{2.28} \label{eq:2.28}
\end{equation}
\begin{equation}
\tilde{T} =  {\frac{1}{\sqrt{\mathcal{V}}}} \Bigg( \frac{768\pi^2 C^2 x^4 + 256\pi^2 C^2 x^2 - 256\pi^2 C^2 \tilde{\mathcal{A}}xy - \tilde{Q}^2}{1024\pi^3 C^2 x^2 (x + 2\tilde{\mathcal{A}}y)} \Bigg),
\tag{2.29} \label{eq:2.29}
\end{equation}
\begin{equation}
\tilde{\Phi}  = {\frac{1}{\sqrt{\mathcal{V}}}} \bigg( \frac{\tilde{Q}}{16\pi C x} \bigg),
\tag{2.30} \label{eq:2.30}
\end{equation}
\begin{equation}
p =  \frac{E}{2\mathcal{V}},
\tag{2.31} \label{eq:2.31}
\end{equation}
\begin{equation}
\mu = \frac{256\pi^2 C^2 x^4 + 256\pi^2 C^2 x^2 + 256\pi^2 C^2 \tilde{\mathcal{A}}xy -  \tilde{Q}^2}{32\pi C^2 x \sqrt{\mathcal{V}}},
\tag{2.32} \label{eq:2.32}
\end{equation}
\begin{equation}
\tilde{\alpha} = \tilde{\mathcal{A}}\ell^2xy.
\tag{2.33} \label{eq:2.33}
\end{equation}
From \eqref{eq:2.15} and \eqref{eq:2.26}, we can easily see that the expression for $y$ depends on $\tilde\alpha$, which is undetermined even $\tilde{\mathcal{A}}$ is fixed. Thus there are two intermediate variables that cannot be determined. Even after eliminating the intermediate variables $x$, the relationships between the thermodynamic quantities are still not fully determined. This is the reason for the unusual behavior of the free energy, as mentioned in Section \ref{sec:3}. 

One might suggest that $G_N$ should be held to be a fixed constant for the boundary CFT, and only varied in the mixed thermodynamics. In fact, ref. ~\cite{Qu:2022nrt}, \cite{Sadeghi:2024ish} and \cite{Yang:2024krx} take the conventional approach of fixing $G_N$.
However, if this proposal were adopted, the quantity 
$\ell^2=16\pi G_{N}C$ would be constant in the ensemble under consideration, which in turn would make the intermediate variable $x$ a constant as well, as seen in equation \eqref{eq:3.4}. This is an outcome we wish to avoid. Of course, one could also treat $G$ as a invariant constant, but this would require one to assign a specific value to $G$ in his calculations, which is an arbitrary choice. This is analogous to how we treat $y$ as a constant here, where its specific value is also arbitrarily chosen. The similarity between these two approaches may hint at the fundamental nature of the results presented in this paper.

\section{Holographic CFT phase transition and criticality in ensemble at fixed $(C, \mathcal{V}, \tilde{Q}, \tilde{\mathcal{A}})$ }
\label{sec:3}

In the ensemble we fix $(C, \mathcal{V}, \tilde{Q}, \tilde{\mathcal{A}})$. The thermodynamic potential in this ensemble is
\begin{equation}
G=E-\tilde{T}\tilde{S}-\tilde{\mathcal{A}}\tilde{\alpha}=\tilde{\Phi}\tilde{Q}+\mu C
\tag{3.1} \label{eq:3.1}
\end{equation}
And the differential of $G$ satisfies
\begin{equation}
\begin{split}
dG &= dE - \tilde{T} d\tilde{S} - \tilde{S} d\tilde{T} -\tilde{\mathcal{A}}d\tilde{\alpha} -\tilde{{\alpha}}d\tilde{\mathcal{A}}\\&= -\tilde{S} d\tilde{T} + \mu dC  - p d\mathcal{V} + \tilde{\Phi} d\tilde{Q}  - \tilde{\alpha}d\tilde{\mathcal{A}}.
\end{split}
\tag{3.2} \label{eq:3.2}
\end{equation}
Therefore, $G$ is stationary at fixed $(\tilde{T},C, \mathcal{V}, \tilde{Q}, \tilde{\mathcal{A}})$. 

\subsection{$d=4$}

Before presenting the expression for the free energy, we wish to engage in a more subtle discussion regarding the thermodynamic conjugate quantities pair $(\tilde{\mathcal{A}}, \tilde{\alpha})$.

From \eqref{eq:2.15} and \eqref{eq:2.26}, we find that $\tilde{\alpha}$ always appear in the expression of $y$ even if $\tilde{A}$ is fixed. It means that in the expressions of $G$ the number of intermediate variables is $2$ ($x$ and $\tilde{\alpha}(y)$). In order to reduce the number of intermediate variables, it is necessary to impose additional constraints. We propose the following constraint when $d=4$:
\begin{equation}
y=\frac{\sqrt{\mathcal{V}}}{8 \pi C}\tilde{\alpha}=\text{constant}.
\tag{3.3} \label{eq:3.3}
\end{equation}
\eqref{eq:3.3} shows that when $C$ and $\mathcal{V}$ are fixed, $\tilde{\alpha}$ is also fixed. However, there is a factor difference for $\tilde{\alpha}$ depending on the value of the constant in \eqref{eq:3.3}. By this method, we achieve our goal of reducing the number of intermediate variables. It is worth noting that the value of $\tilde{\alpha}$ remains independent of $\tilde{\mathcal{A}}$.

Of course, there is another obvious method of constraint. According to \eqref{eq:2.26} and \eqref{eq:2.33}, we can get
\begin{equation}
\frac{1}{\ell^2}=\frac{x\sqrt{\mathcal{V}}}{8 \pi C}\tilde{\mathcal{A}}.
\tag{3.4} \label{eq:3.4}
\end{equation}
Once we fix $(C, \mathcal{V}, \tilde{Q}, \tilde{\mathcal{A}})$, $\ell$
varies with $x$ and this situation can be repaired by fixing $\ell$. 
If $\ell$ is fixed, then $x$ is fixed. Thus the number of intermediate variable is reduced to $1$ leaving only $\tilde\alpha(y)$. Because $\ell= R /\omega$ and $R$ is fixed in this ensemble, so this method equals to fixing $\omega$. This problem stems from our definition of thermodynamic quantities in CFT, and such issues do not arise without introducing higher-order curvature coupling coefficients.

A comparison of the two constraint methods suggests that the first approach is preferable. The second method leads to $\omega=R/\ell$ being fixed, which contradicts the principles of the holographic dictionary \eqref{eq:2.7}\eqref{eq:2.23}. Additionally, fixing $x$ also results in $\tilde{\Phi}$ \eqref{eq:2.30} being fixed, thereby extending the original problem between $\tilde{\mathcal{A}}$ and $\tilde{\alpha}$ to another pair of thermodynamic variables. Although one may question the physical motivation behind the first artificial constraint, we will adopt this constraint for the subsequent study. And for simplicity, we set $y=1$.

Using \eqref{eq:2.26}, \eqref{eq:2.27}, \eqref{eq:2.28}, \eqref{eq:2.29} and \eqref{eq:3.3} with $y=1$, the thermodynamic potential in this ensemble can be obtained:
\begin{equation}
\begin{split}
G &= E - \tilde{T} \tilde{S} -\tilde{\mathcal{A}}\tilde{\alpha}\\&= 
\frac{ \tilde{Q}^2}{256 \pi^2 C^3 \sqrt[3]{\mathcal{V}} x^2} + 
\frac{\tilde{Q}^2 (x^2 + 6\tilde{\mathcal{A}}x)}{384 \pi^2 C x^2 \sqrt[3]{\mathcal{V}} (x^2 + 2\tilde{\mathcal{A}}x)} \\ &+ 
\frac{3C  (x^4 + x^2 + \tilde{\mathcal{A}}x)}{ \sqrt[3]{\mathcal{V}}} \\&- 
\frac{2C  x^2 (2x^2 + 1) \left({\mathcal{V}}\right)^{2/3} (x^2 + 6\tilde{\mathcal{A}}x)}{\mathcal{V} (x^2 + 2\tilde{\mathcal{A}}x)} - \frac{8\pi C \tilde{\mathcal{A}}}{\sqrt{\mathcal{V}}}.
\end{split}
\tag{3.5} \label{eq:3.5}
\end{equation}
The temperature is 
\begin{equation}
\tilde{T} =  \frac{768\pi^2 C^2 x^4 + 256\pi^2 C^2 x^2 - 256\pi^2 C^2 \tilde{\mathcal{A}}x - \tilde{Q}^2}{1024\pi^3 C^2 \sqrt{\mathcal{V}} x (x^2 + 2\tilde{\mathcal{A}}x)}.
\tag{3.6} \label{eq:3.6}
\end{equation}

\subsubsection{Dependence of Free Energy on $\tilde{\mathcal{A}}$}

\begin{figure}
    \centering
    \includegraphics[width=0.85\linewidth]{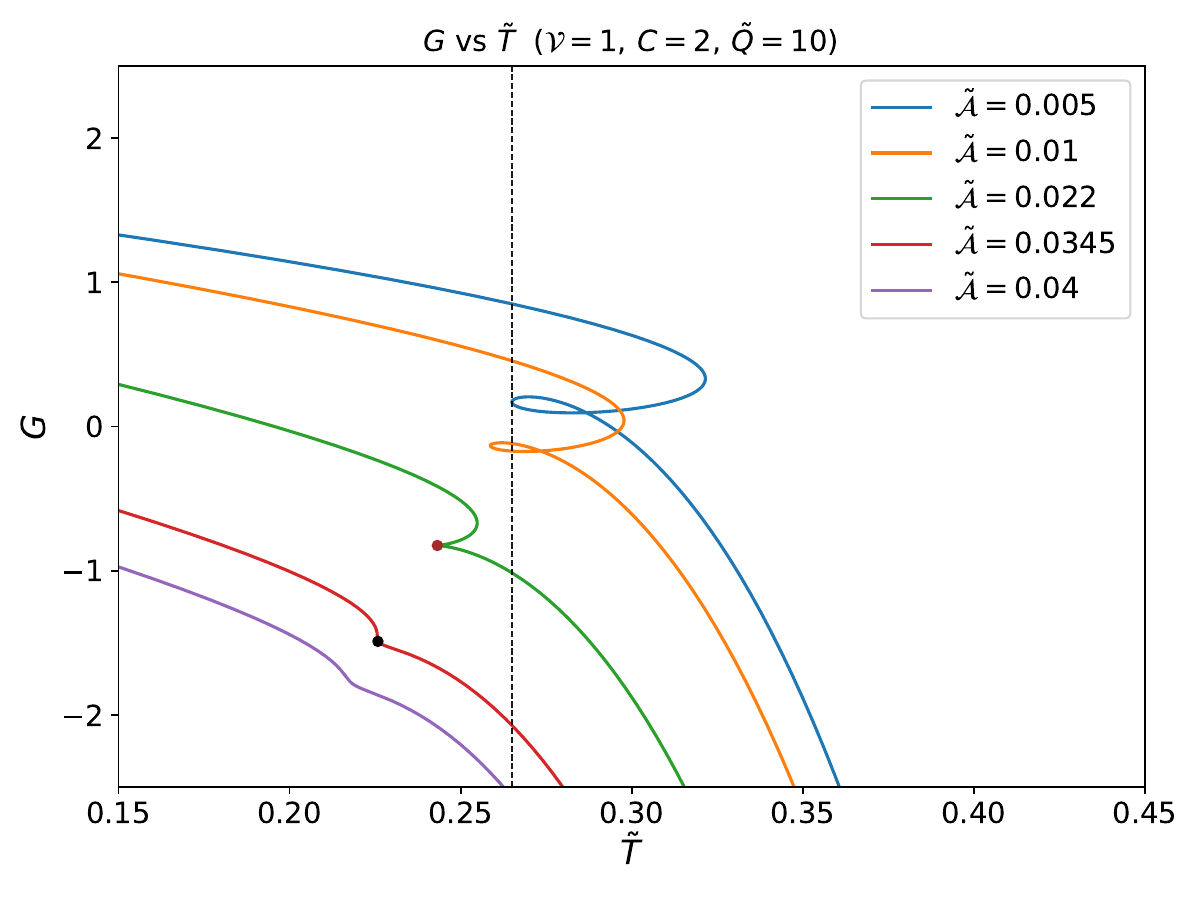}
    \caption{Free energy $G$ vs. temperature $\tilde{T}$ plot in $d = 4$ for the fixed $(C, \mathcal{V}, \tilde{Q}, \tilde{\mathcal{A}})$ ensemble($y=1$). we plot different values of $\tilde{\mathcal{A}}$ for fixed $C$, $\mathcal{V}$ and $\tilde{Q}$, the parameters are $\mathcal{V} = 1$, $C=2$, $\tilde{Q} = 10$ and $\tilde{\mathcal{A}}=0.005, 0.01, 0.022, 0.0345, 0.04$(blue, orange, green, red, purple). }
    \label{fig:1}
\end{figure}

\begin{figure}
    \centering
    \includegraphics[width=0.85\linewidth]{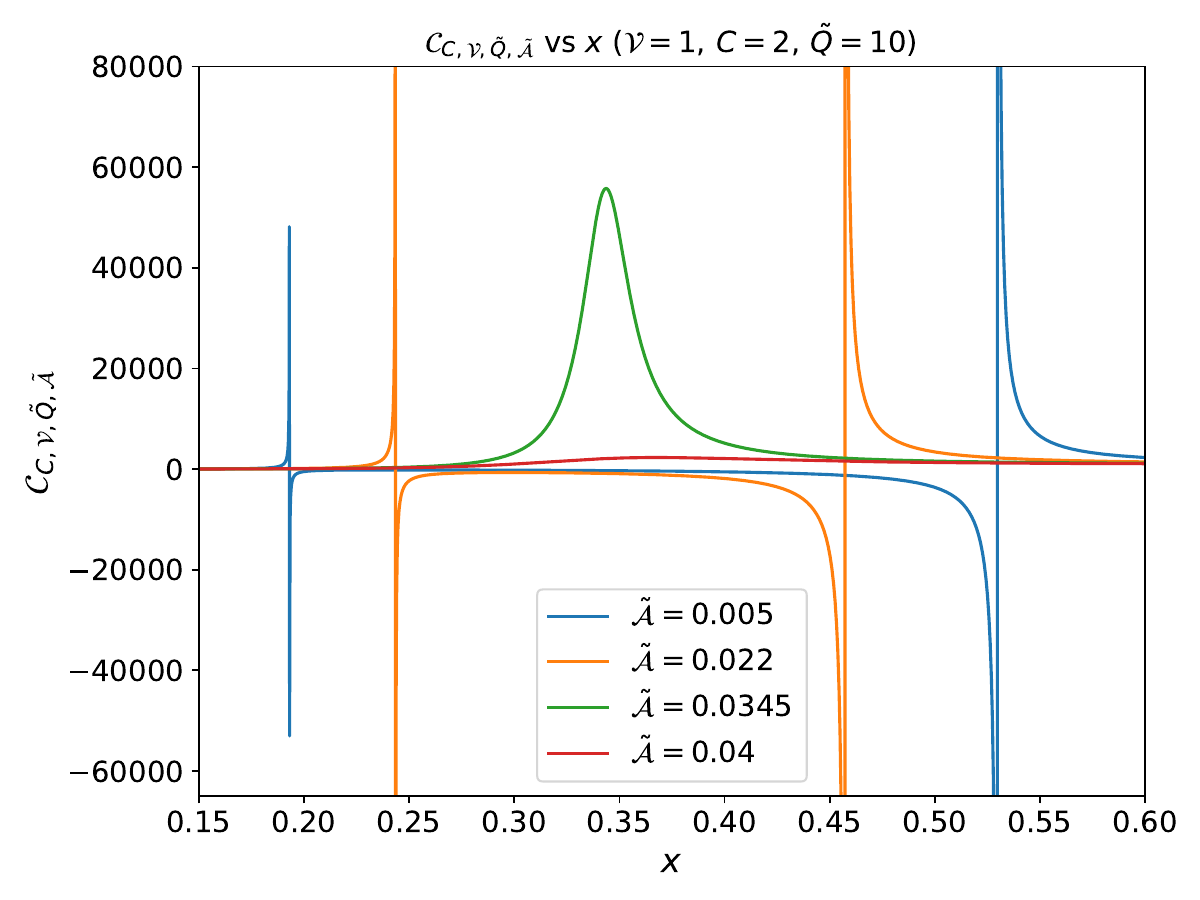}
    \caption{$\mathcal{C}_{C, \mathcal{V}, \tilde{Q},\mathcal{\tilde{A}}}$ vs. $x$ plot in $d=4$ for the fixed $(C, \mathcal{V}, \tilde{Q}, \tilde{\mathcal{A}})$ ensemble($y=1$). we plot different values of $\tilde{\mathcal{A}}$ for fixed $C$, $\mathcal{V}$ and $\tilde{Q}$, the parameters are $\mathcal{V} = 1$, $C=2$, $\tilde{Q} = 10$ and $\tilde{\mathcal{A}}=0.005, 0.022, 0.0345, 0.04$(blue, orange, green, red).}
    \label{fig:2++}
\end{figure}

Let us study how $\tilde{\mathcal{A}}$ influences the behaviour of $G$. At fixed $(C, \mathcal{V}, \tilde{Q})$, \eqref{eq:3.5} and \eqref{eq:3.6} are linked by the mediating variable $x$. The result is displayed in figure \ref{fig:1}.
In figure \ref{fig:1} we show the free energy as a function of the temperature for
$\tilde{\mathcal{A}}< \tilde{\mathcal{A}}^{(1)}_{crit}$ (blue, orange), $\tilde{\mathcal{A}}= \tilde{\mathcal{A}}^{(1)}_{crit}$ (green), $\tilde{\mathcal{A}}= \tilde{\mathcal{A}}^{(2)}_{crit}$ (red) and $\tilde{\mathcal{A}}> \tilde{\mathcal{A}}^{(2)}_{crit}$ (purple) while keeping $C$, $\mathcal{V}$ and $\tilde{Q}$
fixed. And the parameters are $\mathcal{V} = 1$, $C=2$, $\tilde{Q} = 10$. The free energy displays a “loop” shape for $\tilde{\mathcal{A}}< \tilde{\mathcal{A}}^{(1)}_{crit}$
, the loop disappears when $\tilde{\mathcal{A}}= \tilde{\mathcal{A}}^{(1)}_{crit}$.
And a smooth monotonic curve for $\tilde{\mathcal{A}}> \tilde{\mathcal{A}}^{(2)}_{crit}$
. For $\tilde{\mathcal{A}}< \tilde{\mathcal{A}}^{(1)}_{crit}$ (blue, orange) the free energy exhibits a loop presenting two phase transitions. In low temperature a zeroth-order phase transition occurs between one thermodynamically stable branch and one unstable branch. The stability can be assessed by computing the system’s heat capacity:
\begin{flalign}
\begin{split}
&\mathcal{C}_{C, \mathcal{V}, \tilde{Q}, \tilde{\mathcal{A}}} \equiv \tilde{T} \left( \frac{\partial \tilde{S}}{\partial \tilde{T}} \right)_{C, \mathcal{V}, \tilde{Q}, \tilde{\mathcal{A}}}= f_1/f_2,
\\ &f_1 =32 C \pi^2 x (x + 2 \tilde{\mathcal{A}} ) (-\tilde{Q}^2 + 256 C^2 \pi^2 x (x + 3 x^3 - \tilde{\mathcal{A}} ))\\& \times \left(x + \tilde{\mathcal{A}}  + \tilde{\mathcal{A}}  \log\left(\frac{x}{A}\right)\right),
\\&f_2 = \tilde{Q}^2 (3 x + 4 \tilde{\mathcal{A}} ) \\&+ 256 C^2 \pi^2 x (-x^2 + 3 x^4 + 2 \tilde{\mathcal{A}} x  + 12 \tilde{\mathcal{A}} x^3  + 2 \tilde{\mathcal{A}}^2 ).
\end{split}
\tag{3.7} \label{eq:3.7}
\end{flalign}
The numerical result is displayed in figure \ref{fig:2++}. The divergence point of the heat capacity connects the stable and unstable branches.
The two divergence points correspond to the rightmost point of the upper branch and the leftmost point of the middle loop in figure \ref{fig:1}, respectively.

To be specific, as the temperature increases from zero, the free energy suddenly jumps from the upper branch to the lower part of the loop on the middle branch at $\tilde{T}_1$. The upper part of the loop corresponds to a thermodynamically stable phase while the lower part of the loop corresponds to a thermodynamically unstable phase. As can be seen from the blue curve, upon reaching a certain temperature $\tilde{T}_1$ (marked by a vertical black dashed line), the free energy abruptly jumps to the value at the leftmost point of the loop. As the temperature continues to increase, the system remains in a unstable phase until the self-intersection point of the loop. At this self-intersection point, a first-order phase transition occurs between the unstable and stable phases as what we have discussed in previous subsection. And for $\tilde{\mathcal{A}}<\tilde{\mathcal{A}}^{(1)}_{crit}$, increasing $\tilde{\mathcal{A}}$ lowers the temperatures for both zeroth- and first-order phase transitions.
For $\tilde{\mathcal{A}}= \tilde{\mathcal{A}}^{(1)}_{crit}$ (green) there is only a zeroth-order phase transition. The brown dot on the green curve is the critical point where the first-order phase transition disappears. And the shape of the free energy curve (green line) is very similar to that shown in figure 17 of ref. \cite{panigrahi2025}, both exhibiting an unstable state where the free energy folds back with temperature. For $\tilde{\mathcal{A}}= \tilde{\mathcal{A}}^{(2)}_{crit}$ (red), zero-order phase transition also disappears. The black dot on the red curve is the critical point where the zeroth-order phase transition disappears. For $\tilde{\mathcal{A}}> \tilde{\mathcal{A}}^{(2)}_{crit}$ (purple), there is no more zeroth- and first-order transition. However, for certain ranges of $\tilde{\mathcal{A}}> \tilde{\mathcal{A}}^{(2)}_{crit}$ where its value is not sufficiently large, two transitions still occur as the temperature increases, where the system changes from a stable phase to an unstable one, and then back to a stable phase. Yet, the preceding discussion relies on the conventional description of free energy criticality. Next, we will reveal the limitations of this traditional formulation of free energy criticality in this case.

\begin{figure}
    \centering
    \includegraphics[width=0.85\linewidth]{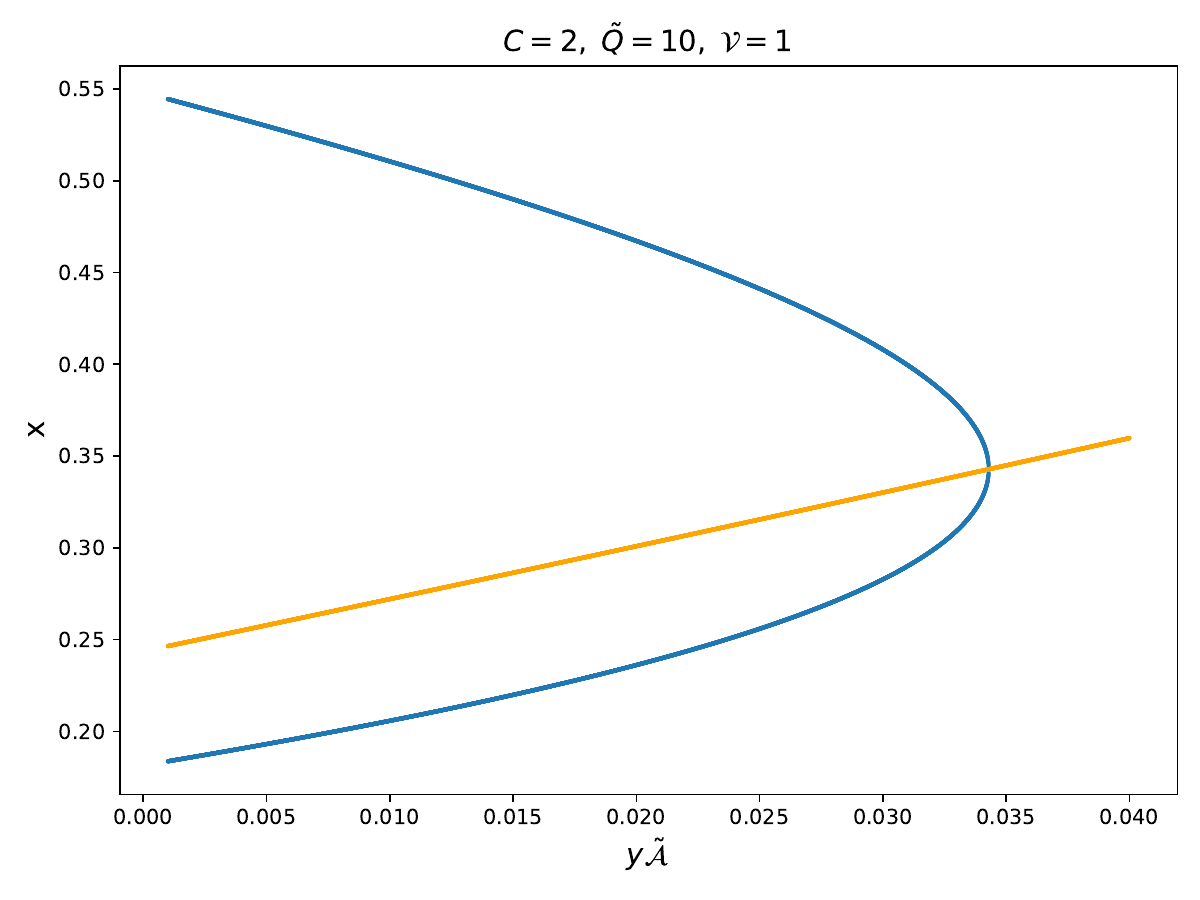}
    \caption{The solutions to \eqref{eq:3.9} and \eqref{eq:3.10} with $\mathcal{V}=1$ , $C=2$ and $\tilde{Q}=10$ for $y\tilde{\mathcal{A}}$ in the range $[0, 0.04]$. The blue dots represent the solution to  \eqref{eq:3.9}, and the orange dots represent the solution to \eqref{eq:3.10}. }
    \label{fig:2}
\end{figure}

\begin{figure}
    \centering
    \includegraphics[width=0.85\linewidth]{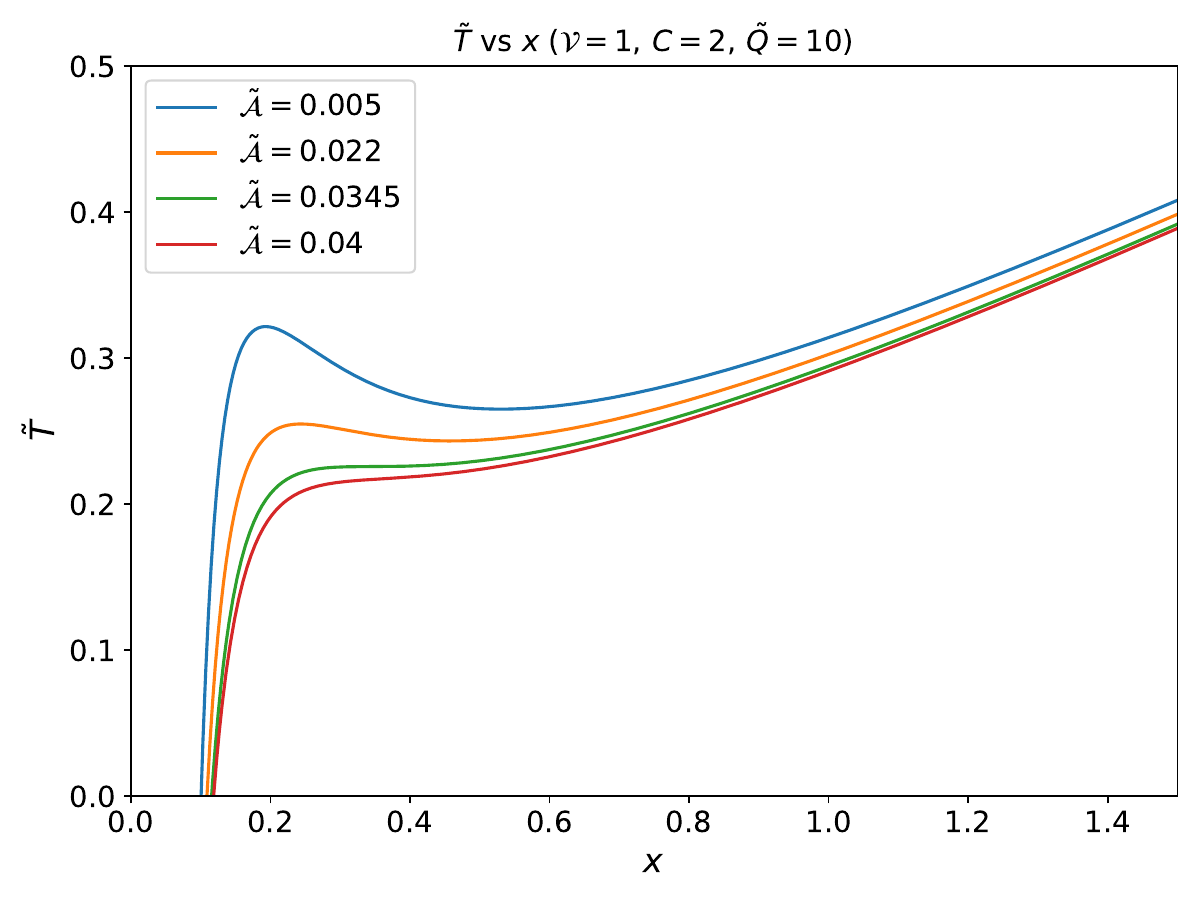}
    \caption{Temperature $\tilde{T}$ vs. $x$ plot in $d=4$ for the fixed $(C, \mathcal{V}, \tilde{Q}, \tilde{\mathcal{A}})$ ensemble($y=1$). we plot different values of $\tilde{\mathcal{A}}$ for fixed $C$, $\mathcal{V}$ and $\tilde{Q}$, the parameters are $\mathcal{V} = 1$, $C=2$, $\tilde{Q} = 10$ and $\tilde{\mathcal{A}}=0.005, 0.022, 0.0345, 0.04$(blue, orange, green, red).}
    \label{fig:2+}
\end{figure}

\begin{figure}
    \centering
    \includegraphics[width=0.85\linewidth]{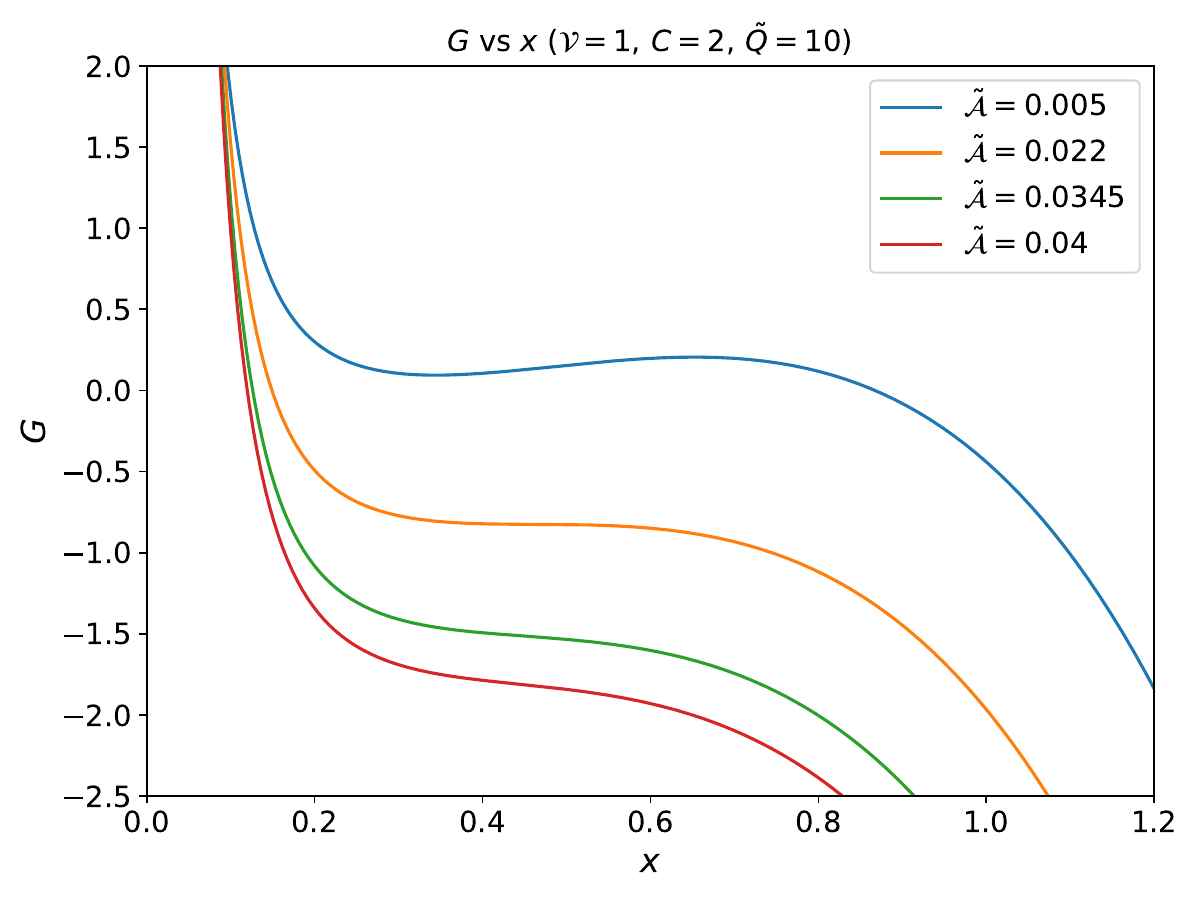}
    \caption{Free energy $G$ vs. $x$ plot in $d=4$ for the fixed $(C, \mathcal{V}, \tilde{Q}, \tilde{\mathcal{A}})$ ensemble($y=1$). we plot different values of $\tilde{\mathcal{A}}$ for fixed $C$, $\mathcal{V}$ and $\tilde{Q}$, the parameters are $\mathcal{V} = 1$, $C=2$, $\tilde{Q} = 10$ and $\tilde{\mathcal{A}}=0.005, 0.022, 0.0345, 0.04$(blue, orange, green, red).}
    \label{fig:2+++}
\end{figure}

In this ensemble with fixed $(C, \mathcal{V}, \tilde{Q}, \tilde{\mathcal{A}})$, the traditional first-order phase transition's critical point is given by the following equation:
\begin{equation}
\left( \frac{\partial \tilde{T}}{\partial x} \right)_{\tilde{Q}, \mathcal{V}, C, \tilde{\mathcal{A}}} = 0, \quad \left( \frac{\partial^2 \tilde{T}}{\partial x^2} \right)_{\tilde{Q}, \mathcal{V}, C, \tilde{\mathcal{A}}} = 0.
\tag{3.8} \label{eq:3.8}
\end{equation}
Using \eqref{eq:3.6}, \eqref{eq:3.8} can be expressed as follows (we have removed the singularity arising from a zero denominator for negative values of $\tilde{\mathcal{A}}$):
\begin{equation}
\begin{split}
768 C^{2} \pi^{2} x^{5} + 3072 y\tilde{\mathcal{A}} C^{2} \pi^{2} x^{4} - 256 C^{2} \pi^{2} x^{3} \\+ 512 y\tilde{\mathcal{A}} C^{2} \pi^{2} x^{2} + (512 y^2\tilde{\mathcal{A}}^{2} C^{2} \pi^{2} + 3 \tilde{Q}^{2}) x \\+ 4y \tilde{\mathcal{A}} \tilde{Q}^{2}=0,
\end{split}
\tag{3.9} \label{eq:3.9}
\end{equation}
\begin{equation}
\begin{split}
(128 C^2 \pi^2 + 1536 C^2 \pi^2 y^2\tilde{\mathcal{A}}^2) x^4 - 384 C^2 \pi^2 y\tilde{\mathcal{A}} x^3 \\+ (-768 C^2 \pi^2 y^2\tilde{\mathcal{A}}^2 - 3 \tilde{Q}^2) x^2 \\+ (-512 C^2 \pi^2 y^3\tilde{\mathcal{A}}^3 - 8 y\tilde{\mathcal{A}} \tilde{Q}^2) x - 6 y^2\tilde{\mathcal{A}}^2 \tilde{Q}^2=0.
\end{split}
\tag{3.10} \label{eq:3.10}
\end{equation}
where we have explicitly written out $y$.
These two equations show that, unlike in the traditional criticality problem, we cannot decouple $\tilde{\mathcal{A}}$ and $y$ at the critical point. In fact, even for the same value of $x$, the two equations yield different values for $\tilde{\mathcal{A}}y$. The solutions to \eqref{eq:3.9} and \eqref{eq:3.10} are shown in Figure \ref{fig:2}. The blue dots represent the solution to  \eqref{eq:3.9}, and the orange dots represent the solution to \eqref{eq:3.10}.
When $y\tilde{\mathcal{A}}$ is in the range of $0$ to $0.04$, the two equations give different roots $x$ for a given $\tilde{\mathcal{A}}y$. The the solution to \eqref{eq:3.9} is divided into an upper and a lower branch, which are connected by the predicted critical point. 
Unexpectedly, the traditional critical point equation \eqref{eq:3.9} and \eqref{eq:3.10} gives $ \tilde{\mathcal{A}}^{(2)}_{crit}=0.0345$ as the critical point of the zeroth-order phase transition in figure \ref{fig:1}, instead of the expected $ \tilde{\mathcal{A}}^{(1)}_{crit}$. This is, however, understandable, as the physics of the equation corresponds to the coincidence of the locations of the rightmost point of the upper branch and the leftmost point of the middle loop.
This result can be seen more clearly in figure \ref{fig:2+}. When $ \tilde{\mathcal{A}}=\tilde{\mathcal{A}}^{(2)}_{crit}=0.0345$, the two extreme points of $\tilde{T}$ merge.
In a sense, this also corroborates the nontrivial dependence of the free energy $G$ on $x$. To identify the subtle loop structure and the critical point of the first-order phase transition, we plotted the free energy as a function of $x$ in figure \ref{fig:2+++}. 
It is evident that for $\tilde{\mathcal{A}}$ less than the critical value $ \tilde{\mathcal{A}}^{(1)}_{crit}$ (blue curve), the free energy’s dependence on $x$ is non-monotonic and exhibits two extrema. When $\tilde{\mathcal{A}}= \tilde{\mathcal{A}}^{(1)}_{crit}$ (orange curve), these two extrema vanish and merge into a plateau. For $\tilde{\mathcal{A}}> \tilde{\mathcal{A}}^{(1)}_{crit}$ (green and red curve), the curve becomes monotonic, and the behavior of the free energy curve as a function of $\tilde{T}$ then depends primarily on the relationship of $\tilde{T}$ with $x$. This explains why the disappearance of the extremum in the $\tilde{T}-x$ curve corresponds to a zeroth-order phase transition. The critical equation for $G$ with respect to $x$, on the other hand, yields the critical point of the first-order phase transition. The proper condition for the critical point $\tilde{\mathcal{A}}^{(1)}_{crit}$ can be expressed as the following equations:
\[\left( \frac{\partial G}{\partial x} \right)_{\tilde{Q}, \mathcal{V}, C, \tilde{\mathcal{A}}} = 0, \quad \left( \frac{\partial^2 G}{\partial x^2} \right)_{\tilde{Q}, \mathcal{V}, C, \tilde{\mathcal{A}}} = 0.\]

Furthermore, this non-monotonic $G-x$ behavior also accounts for the existence of the middle loop structure. But it should be emphasized that this equation cannot be used to obtain the corresponding critical temperature, but only the critical $\tilde{\mathcal{A}}^{(1)}_{crit}$. However, when $\tilde{\mathcal{A}}=\tilde{\mathcal{A}}^{(1)}_{crit}$, the temperature corresponding to the $\tilde{T}-x$ curve's local minimum point is the critical temperature.

Compared to the ensemble with a fixed $\tilde{\alpha}$ discussed in \cite{Sadeghi:2024ish} and \cite{Qu:2022nrt}, the behavior of the free energy becomes more peculiar when both $\tilde{\alpha}$ and $\tilde{\mathcal{A}}$ are fixed. However, when we say that $\tilde{\alpha}$ is fixed here, this fixing is always up to an undetermined constant in \eqref{eq:3.9}. It is the product of this constant $y$ and $\tilde{\mathcal{A}}$ that affects the shape of the free energy curve. For example, the free energy curves are identical for $(y=1, \tilde{\mathcal{A}}=0.01)$ and $(y=2, \tilde{\mathcal{A}}=0.005)$, or for $(y=1, \tilde{\mathcal{A}}=0.01)$ and $(y=0.5, \tilde{\mathcal{A}}=0.02)$. In both cases, they lead to the multiple phase transitions shown in Figure \ref{fig:1}. Therefore, to some extent, it can be argued that the quantity truly affecting the system’s free energy $G$ is actually $\tilde{\mathcal{A}}\tilde{\alpha}$. This is just like in gauge theory, where we always choose a specific gauge to perform our derivations, but the final conclusions are independent of that particular choice. However, when we artificially distinguish an ensemble with a fixed $\tilde{\mathcal{A}}$, the problem arises that $\tilde{\alpha}$ becomes indeterminate and requires the imposition of artificially additional constraints. This issue emerged only after the introduction of the Gauss-Bonnet gravitational coupling coefficient. Given the unique characteristics of this phenomenon figure \ref{fig:1} and figure \ref{fig:2}, we defer the discussion of the analytical results for non-traditional thermodynamic quantities at these critical points in this paper. Indeed, there may be more appropriate approaches for analyzing this ensemble.

\subsubsection{Dependence of Free Energy on $\tilde{Q}$}

Building on the findings from the previous section, where we explored how $\tilde{\mathcal{A}}$ influences the free energy curve, we now turn our attention to the parameter $\tilde{Q}$. Here, we will systematically vary $\tilde{Q}$ while holding $\tilde{\mathcal{A}}$ constant in three distinct ranges: $\tilde{\mathcal{A}} < \tilde{\mathcal{A}}^{(1)}_{crit}$, $\tilde{\mathcal{A}}^{(1)}_{crit}<\tilde{\mathcal{A}} < \tilde{\mathcal{A}}^{(2)}_{crit}$, $\tilde{\mathcal{A}}^{(2)}_{crit}<\tilde{\mathcal{A}}$. Our goal is to uncover more intricate behaviors of the free energy within this ensemble.

\begin{figure}
    \centering
    \includegraphics[width=0.85\linewidth]{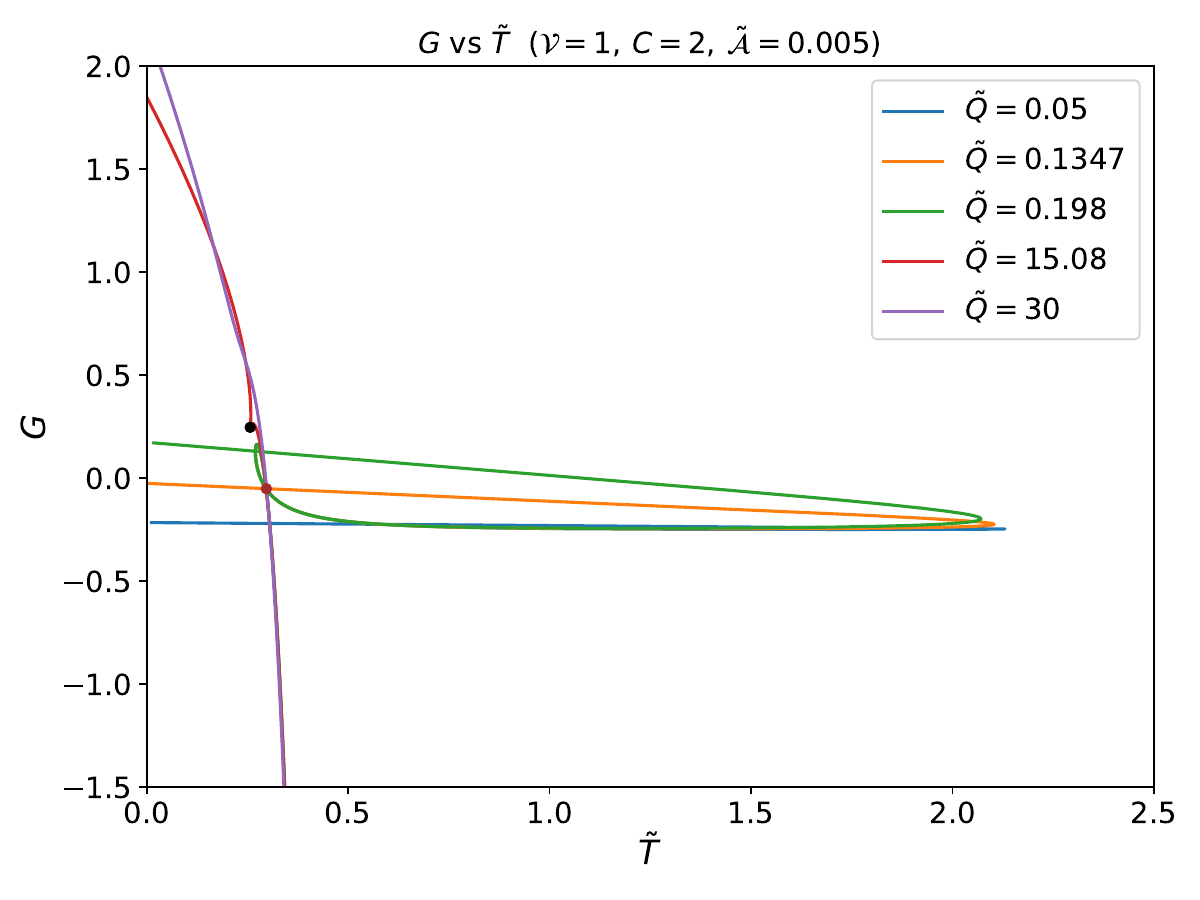}
    \caption{Free energy $G$ vs. temperature $\tilde{T}$ plot in $d = 4$ for the fixed $(C, \mathcal{V}, \tilde{Q}, \tilde{\mathcal{A}})$ ensemble. We plot different values of $\tilde{Q}$ for fixed $C$, $\mathcal{V}$ and $\tilde{\mathcal{A}}$, the parameters are $C = 2$, $\mathcal{V} = 1$, $\tilde{\mathcal{A}}=0.005$ and $\tilde{Q}=0.05, 0.1347, 0.198, 15.08, 30$(blue, orange, green, red, purple).   
    }
    \label{fig:3}
\end{figure}

Firstly, let us fix $\tilde{\mathcal{A}}=0.005 < \tilde{\mathcal{A}}^{(1)}_{crit}$. The result is displayed on figure \ref{fig:3}.
For $\tilde{Q}< \tilde{Q}^{(1)}_{crit}$ (blue) the free energy displays "swallowtail" behaviour and a first-order phase transition occurs between two thermodynamically stable branches.
For $\tilde{Q}= \tilde{Q}^{(1)}_{crit}$(orange) the phase transition point of the free energy coincides with the self-intersection point of the loop in the swallowtail structure. The brown dot marks the critical point on the orange curve.
For $ \tilde{Q}^{(1)}_{crit}<\tilde{Q}< \tilde{Q}^{(2)}_{crit}$ there are two first-order phase transitions. As the temperature increases, a first-order phase transition first occurs between the horizontal branch and the loop. This is a transition from a stable phase to an unstable phase. As the temperature continues to rise, another first-order phase transition occurs between the loop and the vertical branch (self-intersection point of the loop), which is a transition from an unstable phase back to a stable phase.
For $\tilde{Q}= \tilde{Q}^{(2)}_{crit}$ (green) the first first-order phase transition point of the free energy coincides with the leftmost point of the loop in the swallowtail structure. For $\tilde{Q}^{(2)}_{crit}<\tilde{Q}<\tilde{Q}^{(3)}_{crit}$, there are a zeroth-order phase transition and a first-order phase transition. the free energy suddenly jumps from the upper branch to leftmost point of the loop on the middle branch. As the temperature continues to increase, the system remains in a unstable phase until the self-intersection point of the free energy curve. At this self-intersection point, a first-order phase transition occurs between the unstable and stable phases. For $\tilde{Q} ^{(3)}_{crit}\le\tilde{Q}<\tilde{Q}^{(4)}_{crit}$(red), the first-order phase transition disappears, so there is only a zeroth-order phase transition, the critical point is marked by a black point. For $\tilde{Q}^{(4)}_{crit}\le\tilde{Q}$ (purple), there is no more zeroth- and first order phase transition. Furthermore, the high-temperature parts of the free energy curves nearly overlap for all $\tilde{Q}$ values.

\begin{figure}
    \centering
    \includegraphics[width=0.85\linewidth]{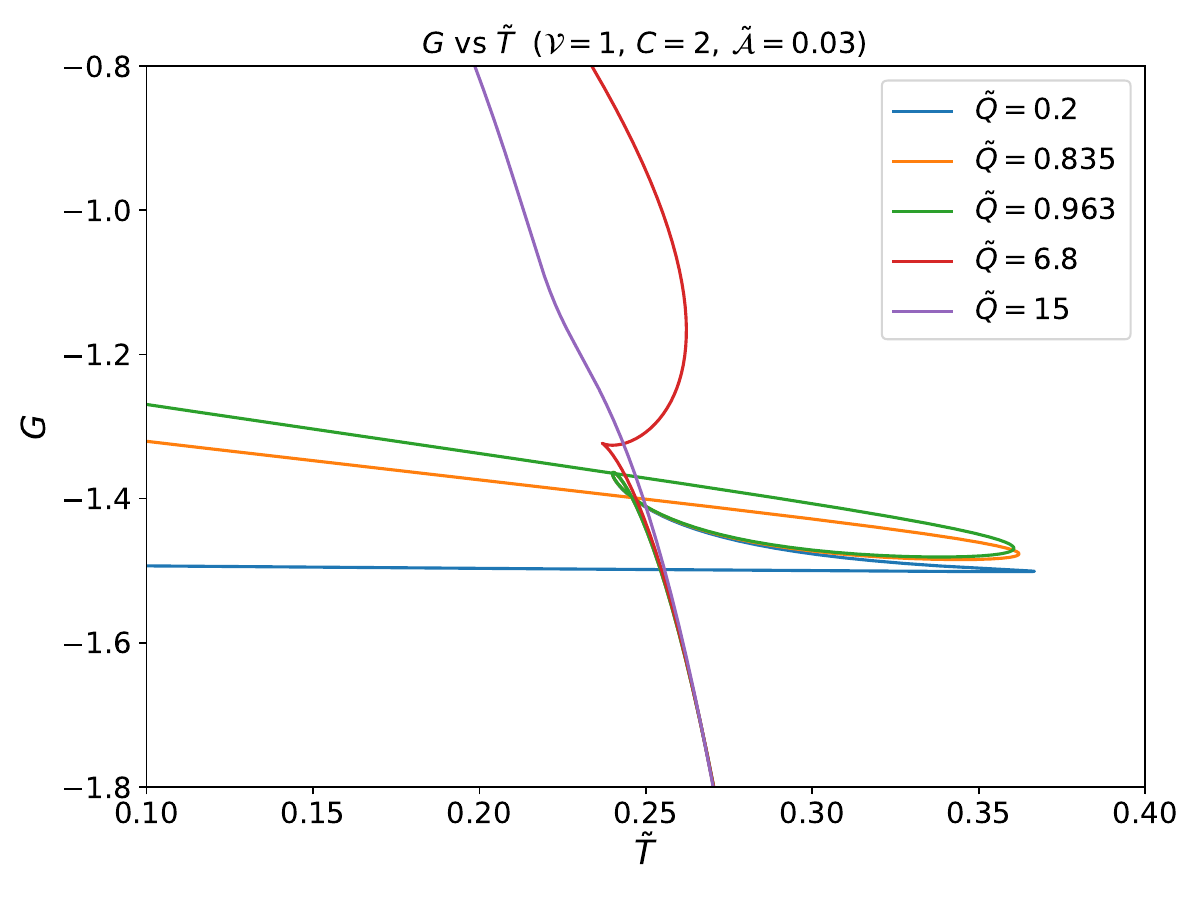}
    \caption{Free energy $G$ vs. temperature $\tilde{T}$ plot in $d = 4$ for the fixed $(C, \mathcal{V}, \tilde{Q}, \tilde{\mathcal{A}})$ ensemble. We plot different values of $\tilde{Q}$ for fixed $C$, $\mathcal{V}$ and $\tilde{\mathcal{A}}$, the parameters are $C = 2$, $\mathcal{V} = 1$, $\tilde{\mathcal{A}}=0.03$ and $\tilde{Q}=0.005, 0.835, 0.963, 6.8, 15$(blue, orange, green, red, purple).   
    }
    \label{fig:4}
\end{figure}

Secondly, let us fix $\tilde{\mathcal{A}}^{(1)}_{crit}<\tilde{\mathcal{A}}=0.03 < \tilde{\mathcal{A}}^{(2)}_{crit}$. The result is displayed on figure \ref{fig:4}.
The free energy behavior is similar to that in figure \ref{fig:3}, likewise featuring critical points and phase transitions. Therefore, we will not elaborate on it here.

\begin{figure}
    \centering
    \includegraphics[width=0.85\linewidth]{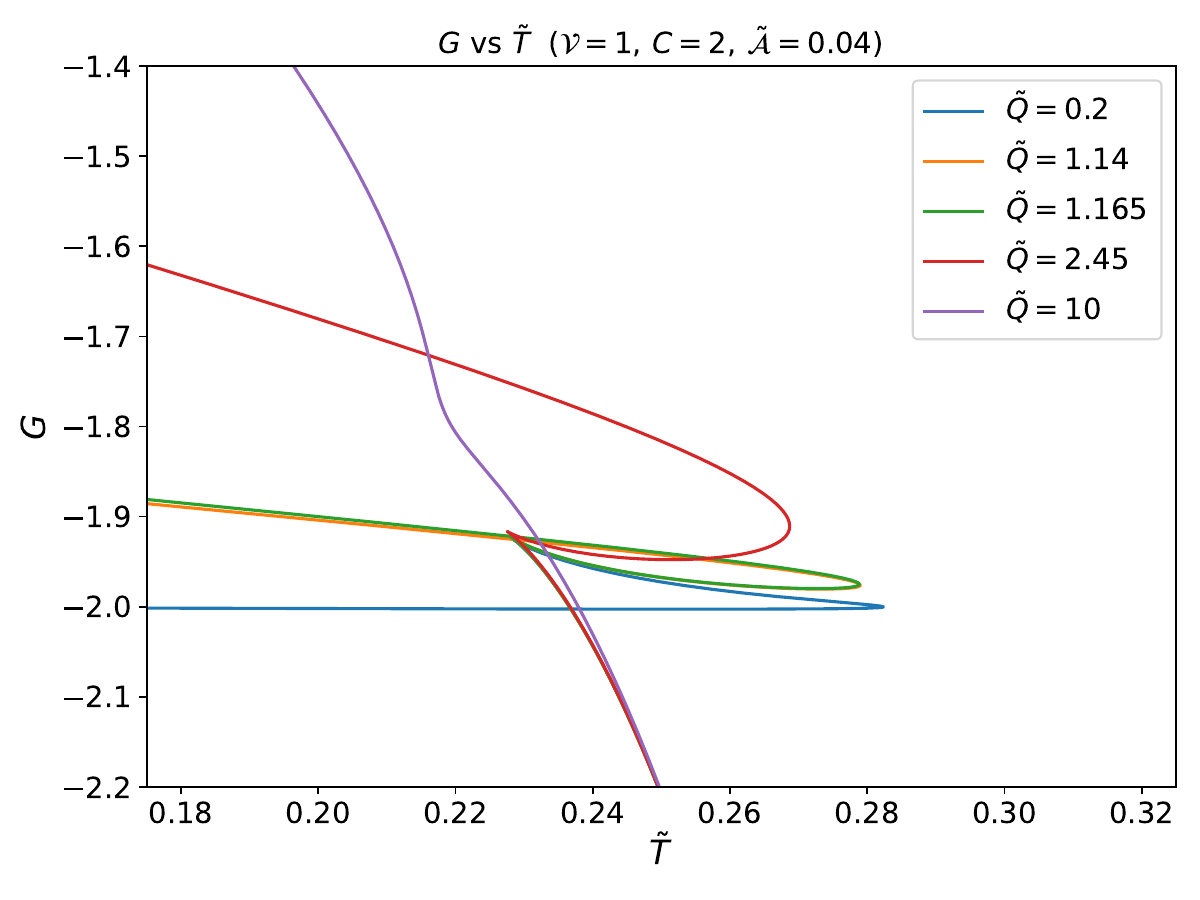}
    \caption{Free energy $G$ vs. temperature $\tilde{T}$ plot in $d = 4$ for the fixed $(C, \mathcal{V}, \tilde{Q}, \tilde{\mathcal{A}})$ ensemble. We plot different values of $\tilde{Q}$ for fixed $C$, $\mathcal{V}$ and $\tilde{\mathcal{A}}$, the parameters are $C = 2$, $\mathcal{V} = 1$, $\tilde{\mathcal{A}}=0.04$ and $\tilde{Q}=0.2, 1.14, 1.165, 2.45, 10$(blue, orange, green, red, purple).   
    }
    \label{fig:5}
\end{figure}

Thirdly, let us fix $\tilde{\mathcal{A}}^{(2)}_{crit}<\tilde{\mathcal{A}}=0.04$. The result is displayed on figure \ref{fig:5}.
The free energy behavior is similar to that in figure \ref{fig:3}, likewise featuring critical points and phase transitions. However, in this case, the leftmost point of the swallowtail loop is also the top point of the swallowtail branch. The remaining features of the free energy curve are consistent with those for the first two values of $\tilde{\mathcal{A}}$ which have been studied.

\begin{figure}
    \centering
    \includegraphics[width=0.85\linewidth]{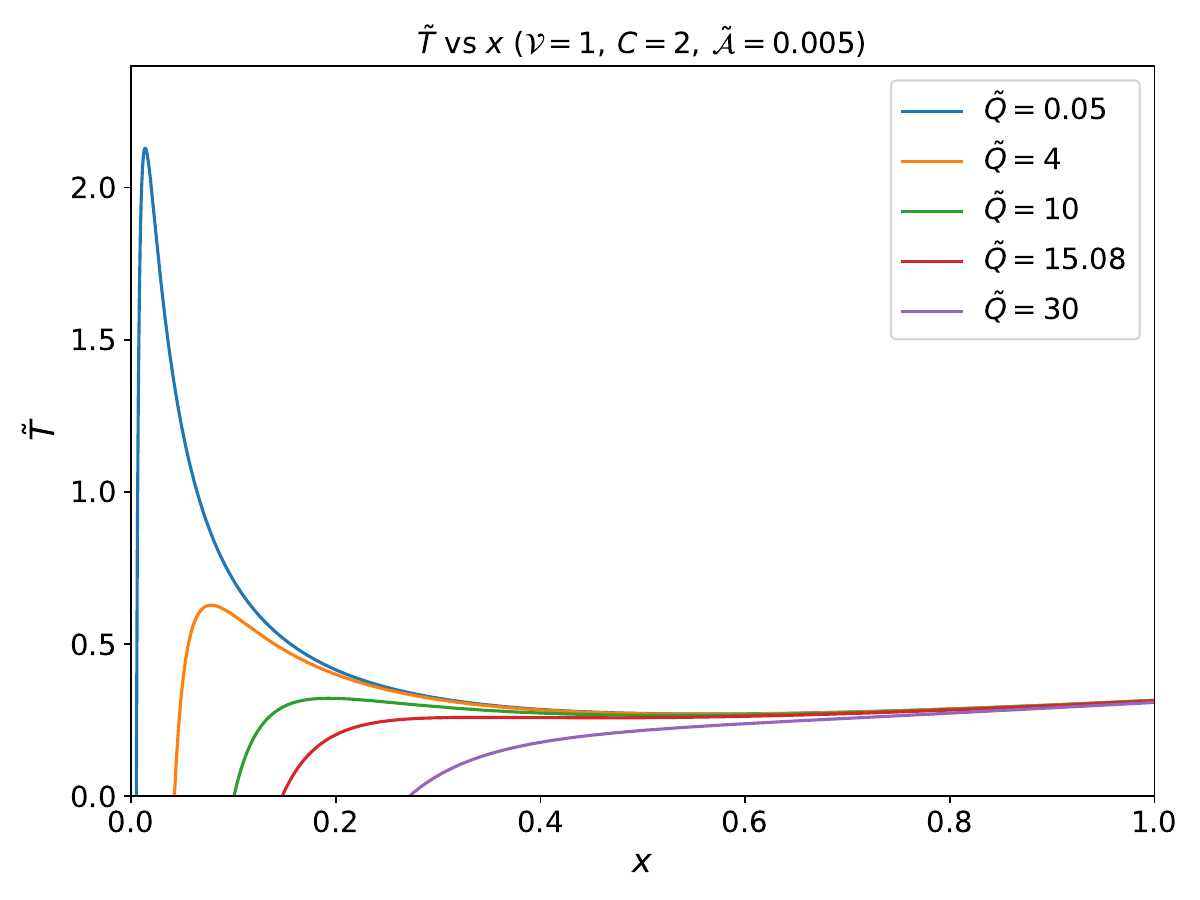}
    \caption{Temperature $\tilde{T}$ vs. $x$ plot in $d=4$ for the fixed $(C, \mathcal{V}, \tilde{Q}, \tilde{\mathcal{A}})$ ensemble. We plot different values of $\tilde{Q}$ for fixed $C$, $\mathcal{V}$ and $\tilde{\mathcal{A}}$, the parameters are $\mathcal{V} = 1$, $C=2$, $\tilde{\mathcal{A}} = 0.005$ and $\tilde{Q}=0.05, 4, 10, 15.08, 30$(blue, orange, green, red, purple).}
    \label{fig:3+}
\end{figure}

\begin{figure}
    \centering
    \includegraphics[width=0.85\linewidth]{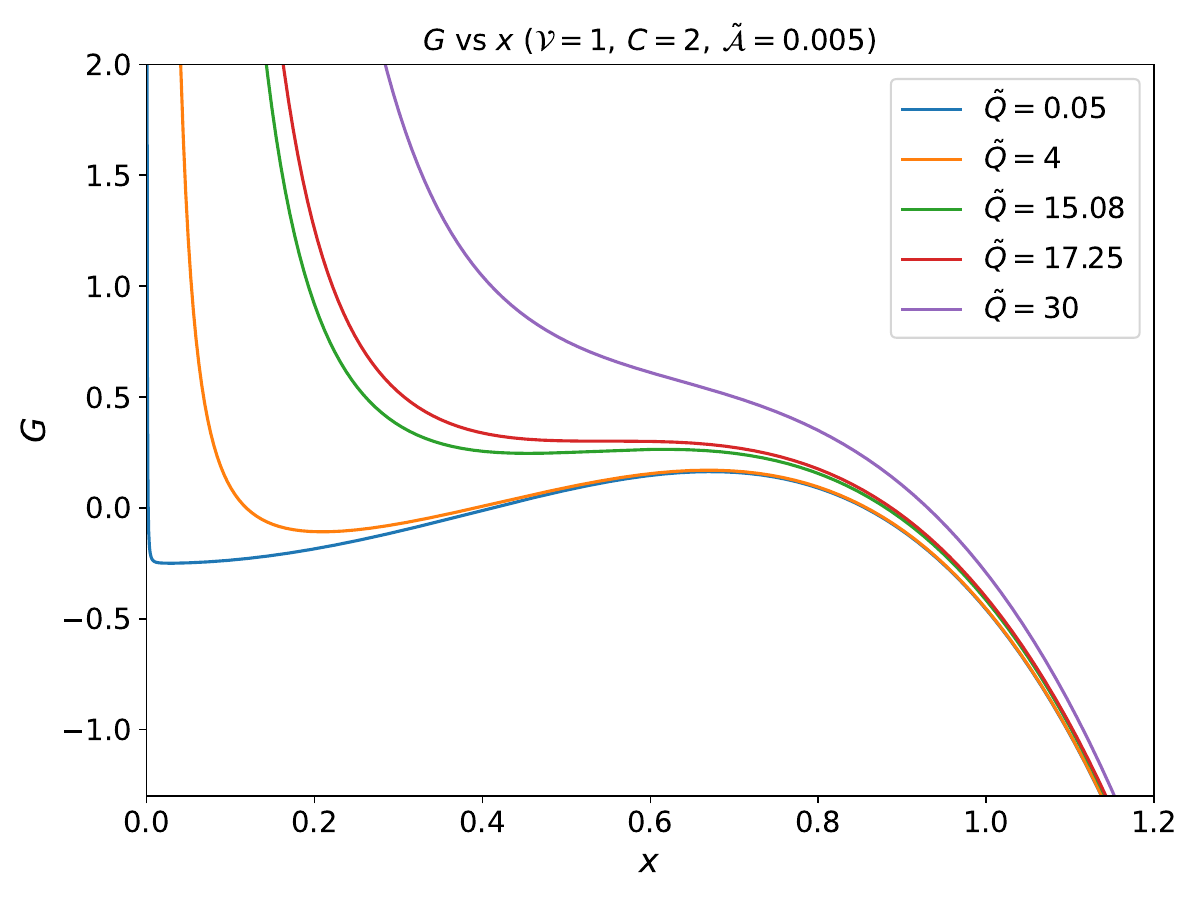}
    \caption{$G$ vs. $x$ plot in $d=4$ for the fixed $(C, \mathcal{V}, \tilde{Q}, \tilde{\mathcal{A}})$ ensemble. We plot different values of $\tilde{Q}$ for fixed $\tilde{\mathcal{A}}$, $\mathcal{V}$ and $\tilde{Q}$, the parameters are $\mathcal{V} = 1$, $C=2$, $\tilde{\mathcal{A}} = 0.005$ and $\tilde{Q}=0.05, 4, 15.08, 17.25,30$(blue, orange, green, red, purple).}
    \label{fig:3++}
\end{figure}

It is not difficult to see that the subtle loop structure also has a significant impact on our classification of phase transitions. In fact, this subtle loop structure also exists in the common swallowtail part, but in those cases, the loop structure does not become the lowest energy state at any temperature. In our findings, the loop structure and its self-intersection point play an important role, as exemplified by the loop structure of the unstable state in a zeroth-order phase transition. What's more, regardless of whether one accepts the loop structure, it is certain that as the value of $\tilde{Q}$ increases from small to large, the type of phase transition changes from a swallow-tail first-order transition to a zeroth-order one. As better illustrated in figure \ref{fig:3+},
the $\tilde{T}-x$ relationship changes from exhibiting two extrema to none as $\tilde{Q}$ varies, closely resembling the characteristics of figure \ref{fig:2+}. The critical value of this transition marks the critical point $\tilde{Q}^{(4)}_{crit}$ (red curve) of the zeroth-order phase transition. We also plotted $G$ as a function of $x$, as shown in figure \ref{fig:3++}. Unlike figure \ref{fig:2++}, the $\tilde{Q}$ value at which the $G-x$ extrema coincide (red curve) does not correspond to any critical behavior in figure \ref{fig:3}. Therefore, we only identified $\tilde{Q}^{(4)}_{crit}$ through the analysis of the free energy curve.

Since analyzing the effect of $\tilde{Q}$ on the free energy demands such a fine-grained approach, we will not address the influence of $C$ or $\mathcal{V}$. And we even have not obtained the critical conditions corresponding to the remaining critical $\tilde{Q}_{crit}$. It is possible that once we find a proper way to comprehend the free energy’s anomalous behavior, all the issues will resolve themselves. 

\subsection{$d=5$}

In order to reduce the number of intermediate variables, it is necessary to impose additional constraints. We propose the following constraint when $d=5$:
\begin{equation}
\quad y \equiv \frac{\mathcal{V}^{1/3}\tilde{\alpha}}{3C}=\text{constant}.
\tag{3.11} \label{eq:3.11}
\end{equation}
After taking $y=1$ and using \eqref{eq:2.16} \eqref{eq:2.17} \eqref{eq:2.18}, the expression of the free energy $F$ is shown as follows:
\begin{equation}
\begin{split}
F &= E - \tilde{T} \tilde{S} -\tilde{\mathcal{A}}\tilde{\alpha} \\&= 
\frac{\tilde{Q}^2}{256 \pi^2 C {\mathcal{V}}^{1/3} x^2}+ 
\frac{\tilde{Q}^2 (x^2 + 6\tilde{\mathcal{A}})}{384 \pi^2 C x^2 {\mathcal{V}}^{1/3} (x^2 + 2\tilde{\mathcal{A}})} \\&+
\frac{3C  (x^4 + x^2 + \tilde{\mathcal{A}})}{{\mathcal{V}}^{1/3}} -
\frac{2C  x^2 (2x^2 + 1) (x^2 + 6\tilde{\mathcal{A}})}{{\mathcal{V}}^{1/3} (x^2 + 2\tilde{\mathcal{A}})} \\&- \frac{3C\tilde{\mathcal{A}}}{{\mathcal{V}}^{1/3}} .
\end{split}
\tag{3.12} \label{eq:3.12}
\end{equation}
And the temperature is
\begin{equation}
\begin{split}
\tilde{T}=  \frac{- \tilde{Q}^2 + 1536 \pi^2 C^2 x^6 + 768 \pi^2 C^2 x^4}{1536 \mathcal{V}^{1/3}\pi^3 C^2 x^3 (x^2 + 2\tilde{\mathcal{A}})}.
\end{split}
\tag{3.13} \label{eq:3.13}
\end{equation}
\begin{figure}
    \centering
    \includegraphics[width=0.85\linewidth]{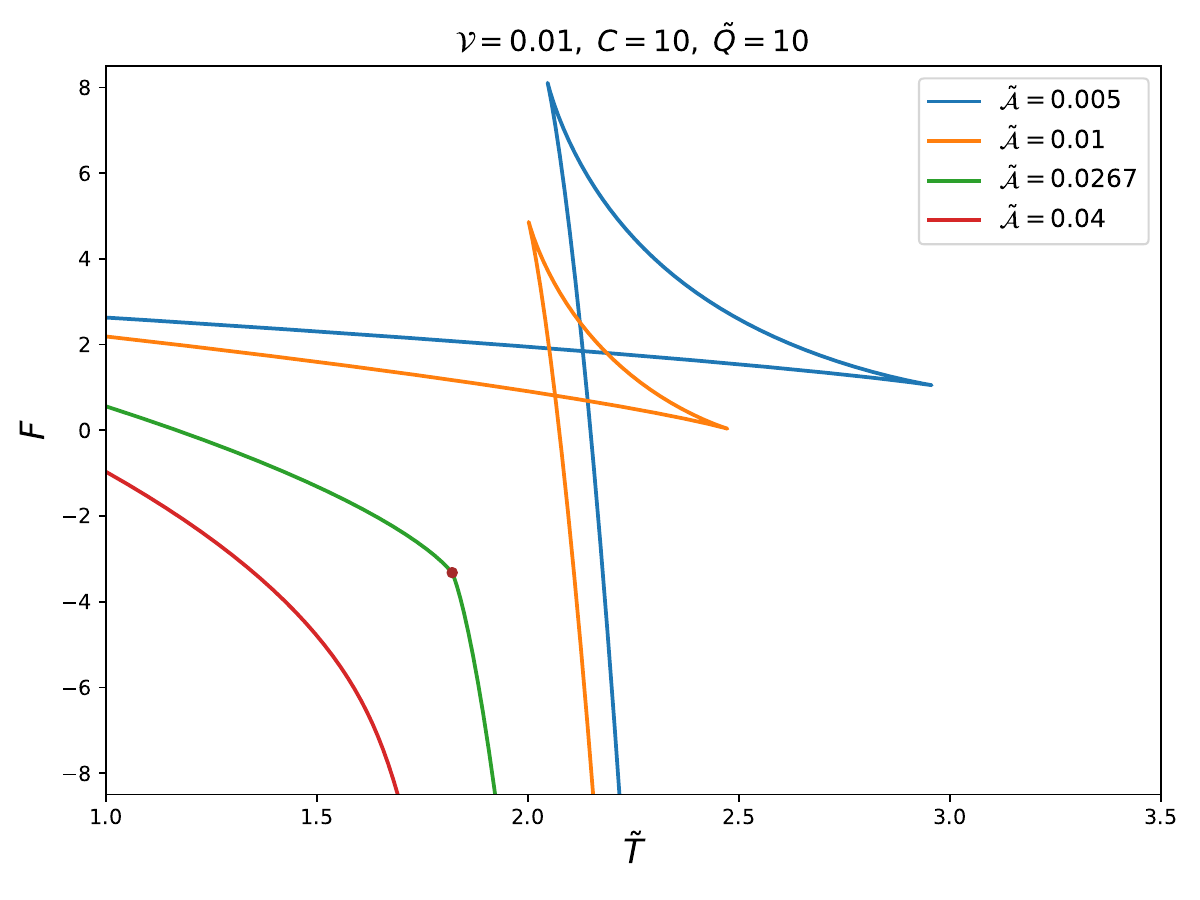}
    \caption{Free energy $G$ vs. temperature $\tilde{T}$ plot in $d = 5$ for the fixed $(C, \mathcal{V}, \tilde{Q}, \tilde{\mathcal{A}})$ ensemble($y=1$). we plot different values of $\tilde{\mathcal{A}}$ for fixed $C$, $\mathcal{V}$ and $\tilde{Q}$, the parameters are $\mathcal{V} = 0.01$, $C=10$, $\tilde{Q} = 10$ and $\tilde{\mathcal{A}}=0.005, 0.01, 0.0267, 0.04$(blue, orange, green, red). The critical point is marked by a brown point.}
    \label{fig:6}
\end{figure}
In figure \ref{fig:6} we show the free energy as a function of the temperature for
$\tilde{\mathcal{A}}< \tilde{\mathcal{A}}_{crit}$ (blue, orange), $\tilde{\mathcal{A}}= \tilde{\mathcal{A}}_{crit}$ (green) and $\tilde{\mathcal{A}}> \tilde{\mathcal{A}}_{crit}$ (red), while keeping $C$, $\mathcal{V}$ and $\tilde{Q}$
fixed. And the parameters are $\mathcal{V} = 0.01$, $C=10$, $\tilde{Q} = 10$. The free energy displays a “swallowtail” shape for $\tilde{\mathcal{A}}< \tilde{\mathcal{A}}_{crit}$
, a kink when $\tilde{\mathcal{A}}= \tilde{\mathcal{A}}_{crit}$,
and a smooth monotonic curve for $\tilde{\mathcal{A}}> \tilde{\mathcal{A}}_{crit}$. For each of the curves, starting from the
point on the curve where $\tilde{T} = 0$, the value of $x$ along the curves increases as $\tilde{T}$ increases.
From the formula \eqref{eq:2.17}
for the CFT entropy, we see that black holes with small
$x \equiv r_+/\ell$ are dual to CFT thermal states with small $\tilde{S}/C$, which are states with low entropy per degree of freedom.
On the swallowtail curve (e.g. blue), this low-entropy state is the only available
state near $\tilde{T} = 0$ on this curve and thus has initially the lowest free energy $F$. It continues to have the lowest free energy as $\tilde{T}$ increases until the self-intersection point of the curve. Beyond this point, the CFT state with high entropy per degree of freedom, corresponding to
large $x$ black holes, lying along the “vertical” branch of the curve, becomes the state with
lowest free energy $F$ and hence dominates the canonical ensemble. A first-order phase
transition thus takes place between low- and high-entropy states at the self-intersection
temperature for each value of $\tilde{\mathcal{A}}< \tilde{\mathcal{A}}_{crit}$. However, when $\tilde{\mathcal{A}}= \tilde{\mathcal{A}}_{crit}$ there only a second-order phase transition between low- and high-entropy states at the critical point. As we increase $\tilde{\mathcal{A}}$, the temperature at which the first-order phase transition
occurs decreases. As shown in figure \ref{fig:6}, when $d = 5$, the behavior of the free energy is similar to that of the ensemble at fixed $(C, \mathcal{V}, \tilde{Q}, \tilde{\alpha})$ \cite{Sadeghi:2024ish}. This is because Equation \ref{eq:2.22} differs from Equation \ref{eq:2.33} in that it does not contain the intermediate variable $x$. Consequently, once $y$ is set to $1$, the $\tilde{\mathcal{A}}\tilde{\alpha}$ term in the free energy $F$ manifests only as an additional constant compared to the free energy free energy in the ensemble at fixed $(C, \mathcal{V}, \tilde{Q}, \tilde{\alpha})$ \cite{Sadeghi:2024ish}.

\begin{figure}
    \centering
    \includegraphics[width=0.85\linewidth]{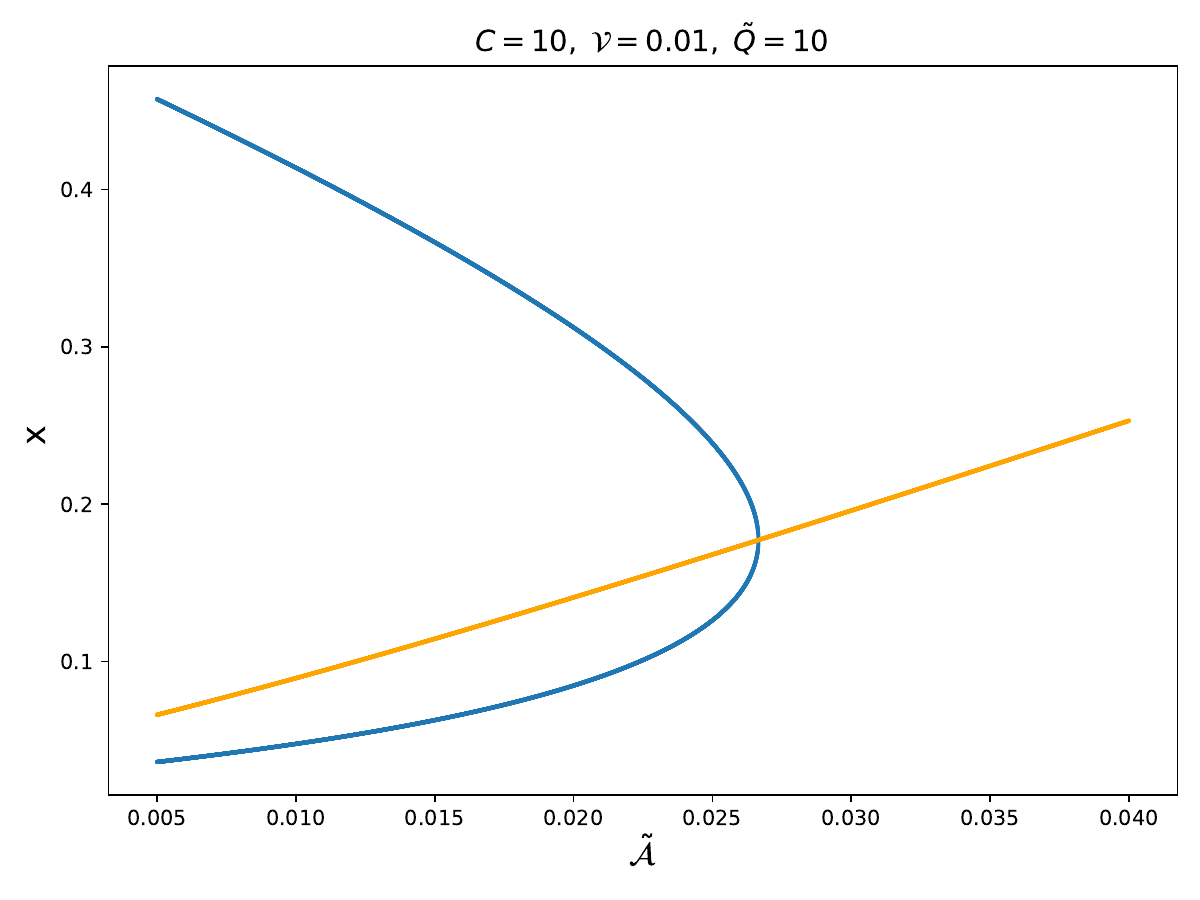}
    \caption{The solutions to \eqref{eq:3.14} and \eqref{eq:3.15} with $\mathcal{V}=0.01$ , $C=10$ and $\tilde{Q}=10$. The blue dots represent the solution to  \eqref{eq:3.14}, and the orange  dots represent the solution to \eqref{eq:3.15}. }
    \label{fig:7}
\end{figure}

Using \eqref{eq:3.8} and \eqref{eq:3.13} we can also get the equations for critical point(we have removed the singularity arising from a zero denominator for negative values of $\tilde{\mathcal{A}}$):
\begin{equation}
\begin{split}
&1536 C^{2} \pi^{2} x^{8} + 768 (12 \tilde{\mathcal{A}} - 1) C^{2} \pi^{2} x^{6} \\&+ 1536 \tilde{\mathcal{A}} C^{2} \pi^{2} x^{4} + 5 \tilde{Q}^{2} x^{2} + 6 \tilde{\mathcal{A}} \tilde{Q}^{2}=0,
\end{split}
\tag{3.14} \label{eq:3.14}
\end{equation}
and
\begin{equation}
\begin{split}
&-768 (4\tilde{\mathcal{A}} - 1) \tilde{Q}^{2} \pi^{2} x^{8} + 4608 \tilde{\mathcal{A}} (4\tilde{\mathcal{A}} - 1) \tilde{Q}^{2} \pi^{2} x^{6} \\&- 15 \tilde{Q}^{2} x^{4} - 34 \tilde{\mathcal{A}} \tilde{Q}^{2} x^{2} - 24 \tilde{\mathcal{A}}^{2} \tilde{Q}^{2}=0.
\end{split}
\tag{3.15} \label{eq:3.15}
\end{equation}
The solution is displayed on figure \ref{fig:7}.
Figure \ref{fig:7} shows that for $\tilde{\mathcal{A}}$ in the range of $0.005$ to $0.04$, the two equations share only one common root for given $\tilde{\mathcal{A}}$. Moreover, equation \eqref{eq:3.14} has two branches, an upper and a lower one. The intersection of these two branches corresponds precisely to the location of the predicted critical point. This behavior closely resembles what is shown in figure \ref{fig:2}. But it gives us the only critical point in \ref{fig:6} which is different from the $d=4$ case. 


We will conclude our investigation of the free energy behavior for $d=5$ here, for the reasons stated at the end of the previous subsection. This paper merely raises this issue and provides a preliminary analysis of this series of problems from the perspective of traditional free energy studies.

\section{Relationships of conjugate thermodynamic pairs}
\label{sec:4}

Since our current understanding of the phase transition behavior is limited to numerical results, the theoretical basis we can rely on is confined to only a few physical laws. Anyway, in this section, we will investigate the relationships between other conjugate thermodynamic pairs (equation of state), such as $C$ and $\mu$, or $p$ and $\mathcal{V}$, or $\tilde{T}$ and $\tilde{S}$ . The aim is to gain a deeper understanding of the CFT's thermodynamic system in the ensemble at fixed $(C, \mathcal{V}, \tilde{Q}, \tilde{\mathcal{A}})$ for $d=4$.  

\subsection{$\mu-C$ relationship}

\begin{figure}
    \centering
    \includegraphics[width=0.85\linewidth]{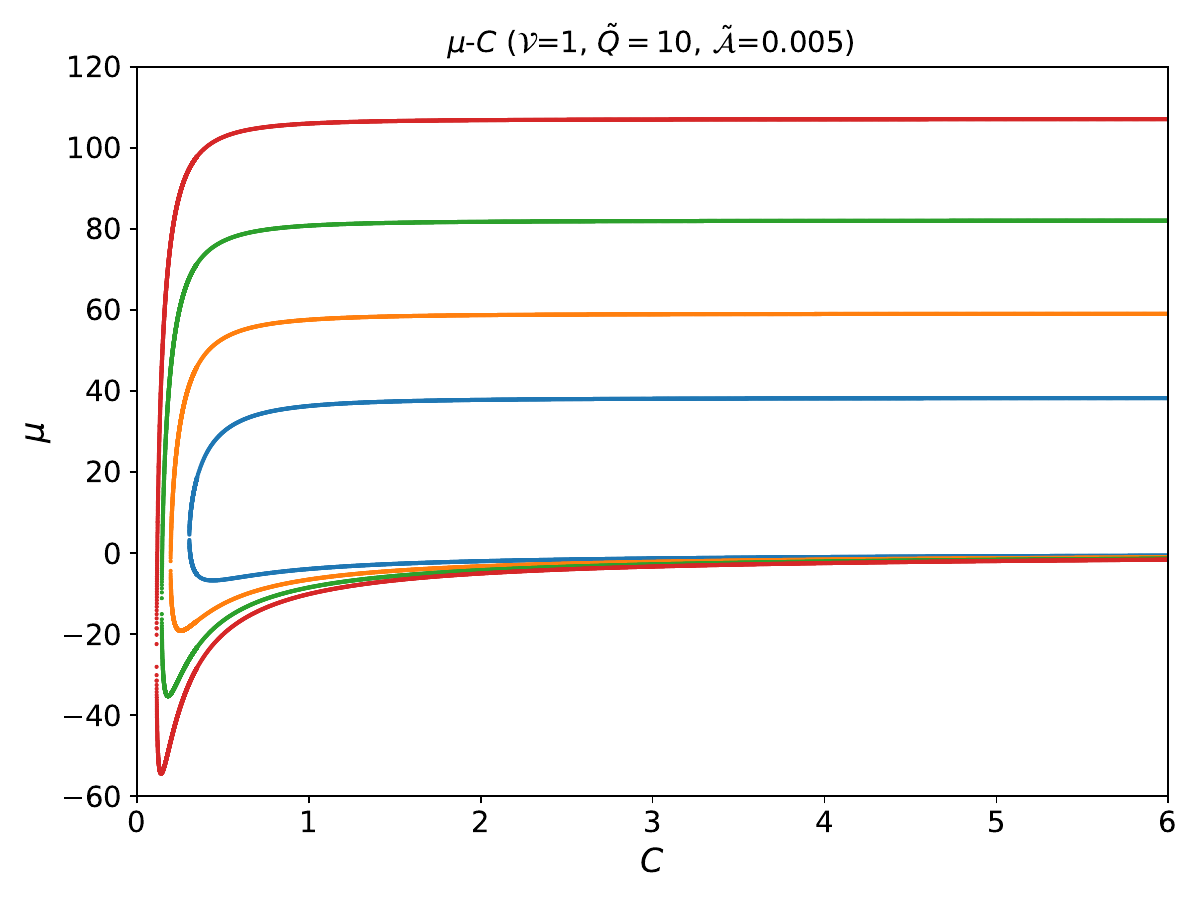}
    \caption{$\mu-C$ isotherms curves for different fixed $\tilde{T}$ ($\tilde{T}=0.1$(blue), $\tilde{T}=0.2$(orange),
    $\tilde{T}=0.3$(green), $\tilde{T}=0.4$(red)) with $\mathcal{V}=1.0$, $\tilde{Q}=10.0$, $\tilde{\mathcal{A}}=0.005$ when $d=4$.}
    \label{fig:8}
\end{figure}

\begin{figure}
    \centering
    \includegraphics[width=0.85\linewidth]{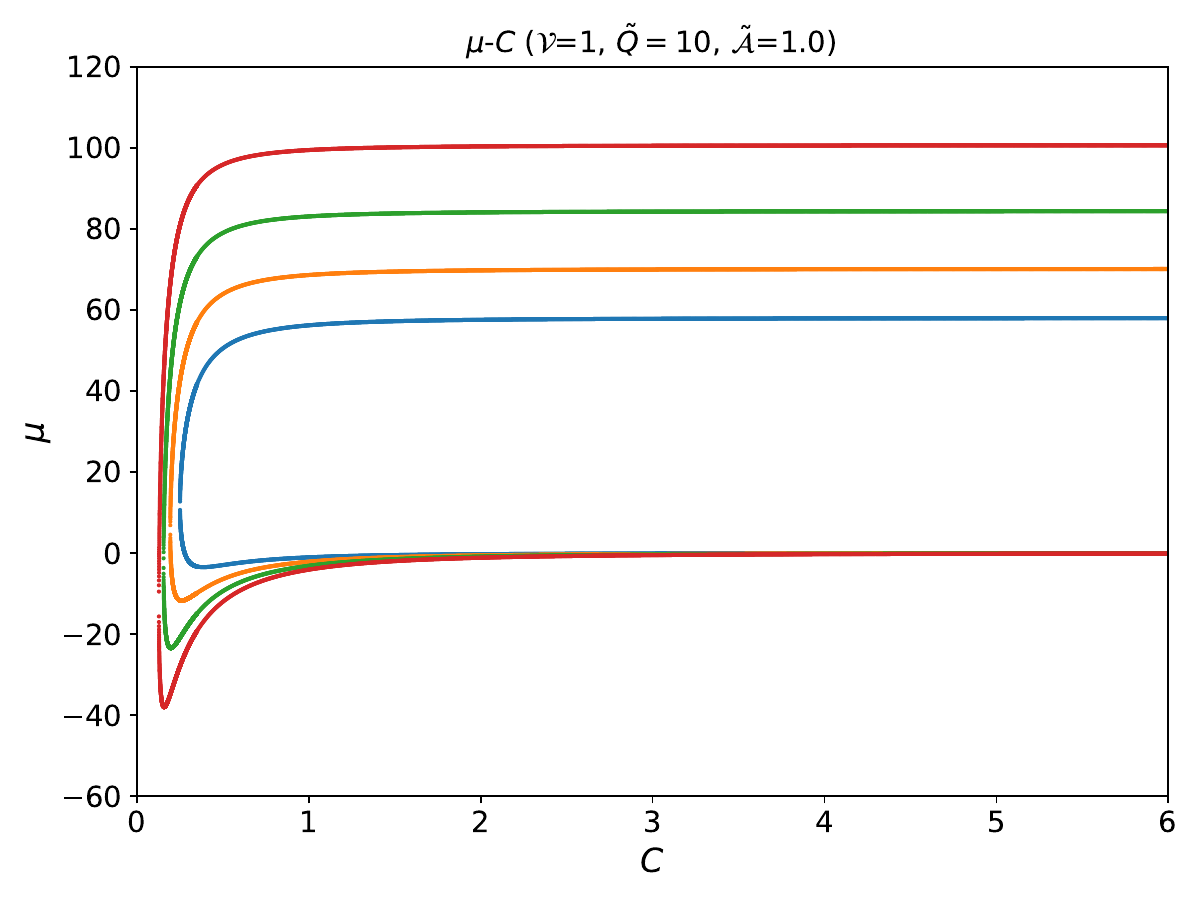}
    \caption{$\mu-C$ isotherms curves for different fixed $\tilde{T}$ ($\tilde{T}=0.1$(blue), $\tilde{T}=0.2$(orange),
    $\tilde{T}=0.3$(green), $\tilde{T}=0.4$(red)) with $\mathcal{V}=1.0$, $\tilde{Q}=10.0$, $\tilde{\mathcal{A}}=0.04$ when $d=4$.}
    \label{fig:9}
\end{figure}

For $d=4$, the chemical potential $\mu$ is a function of $(x,C,\tilde{\mathcal{A}},\mathcal{V})$ as shown in \eqref{eq:2.32}. After setting $y=1$, we can obtain
\begin{equation}
\mu = \frac{256\pi^2 C^2 x^4 + 256\pi^2 C^2 x^2 + 256\pi^2 C^2 \tilde{\mathcal{A}}x -  \tilde{Q}^2}{32\pi C^2 x \sqrt{\mathcal{V}}}.
\tag{4.1} \label{eq:4.1}
\end{equation}
In our study, we still fix $\mathcal{V}=1$ and $\tilde{Q}=10$, but we varies the value of $\tilde{\mathcal{A}}$ and $\tilde{T}$, which means to investigating the effect of $\tilde{\mathcal{A}}$ on the family of $\mu-C$  isotherms. Specifically, for each fixed $\tilde{T}$, we can assign different values to $\tilde{\mathcal{A}}$, solve for $x(C)$ using equation \eqref{eq:3.6}, and then substitute the values of $\tilde{\mathcal{A}}$ and $C$ into equation \eqref{eq:4.1} to obtain different $\mu-C$  curves. 
Results are displayed in figure \ref{fig:8} and \ref{fig:9}, where we only show the positive branch of $x(C)$, because only positive $x$ is physical.
Regardless of the value of $\tilde{\mathcal{A}}$, the family of isotherms exhibits similar characteristics. From the two figures, we can observe that $\mu$ consists of two horizontal branches and one vertical branch. The two horizontal branches indicate that the range of $\mu$ is confined to a closed interval, while the vertical branch suggests the existence of a minimum $C$ value in the isothermal process. The lower horizontal branches of the $\mu-C$ curves almost coincide. For a given value of $\tilde{\mathcal{A}}$, a higher temperature in the isothermal process corresponds to a wider range of $\mu$ and a smaller minimum $C$ value. At a fixed temperature, the range of $\mu$ expands as $\tilde{\mathcal{A}}$ increases at low temperatures, whereas the curves at high temperatures are barely affected. The overall effect is that the low-temperature curves (blue), enveloped by the high-temperature ones, progressively converge towards the high-temperature curves.
This does not exhibit the characteristics of a common van der Waals fluid phase transition, such as Maxwell’s equal-area rule. This is understandable, given that the phase transition behavior of the ensemble under consideration is so complex and puzzling. However, this behavior is very similar to figure 18 in ref \cite{panigrahi2025}.

\subsection{$p-\mathcal{V}$ relationship}

\begin{figure}
    \centering
    \includegraphics[width=0.85\linewidth]{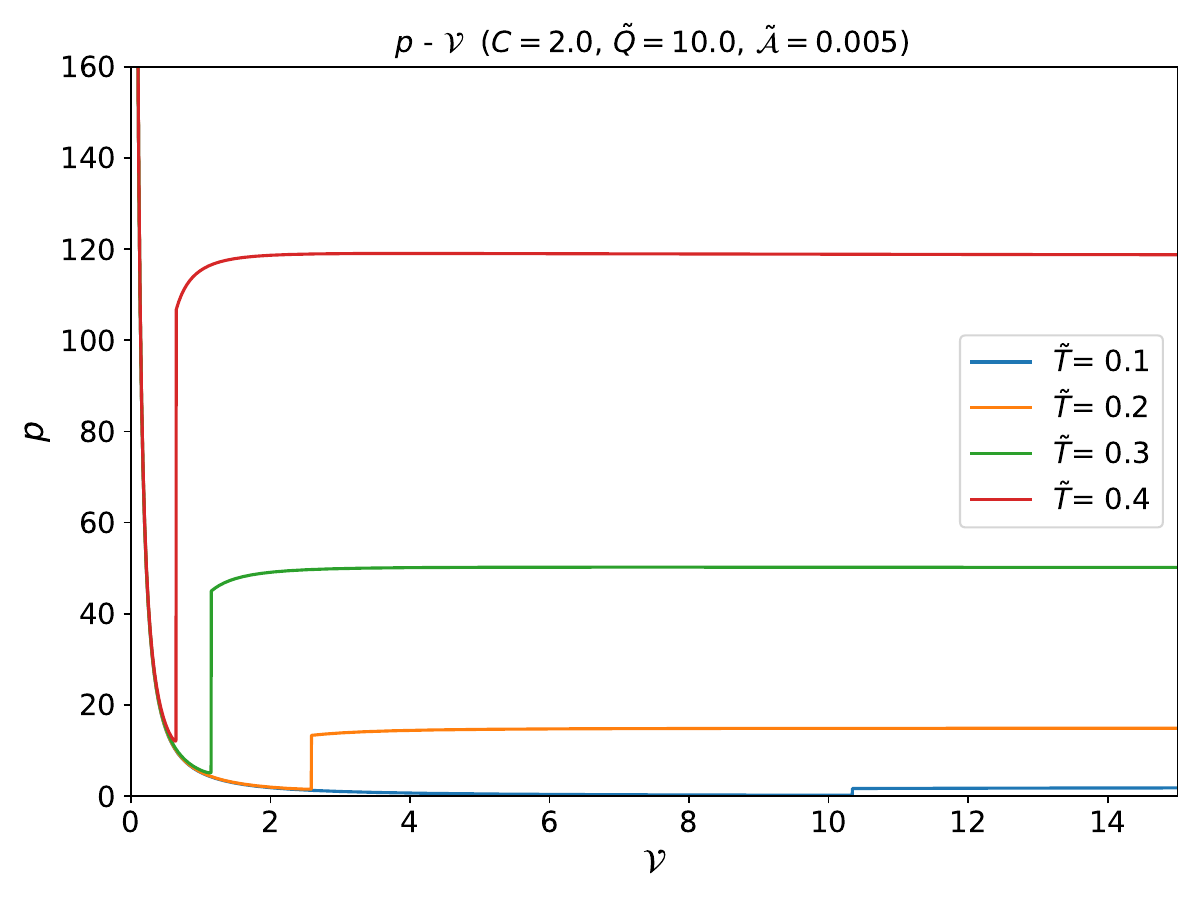}
    \caption{$p-\mathcal{V}$ isotherms curves for different fixed $\tilde{T}$ ($\tilde{T}=0.1$(blue), $\tilde{T}=0.2$(orange),
    $\tilde{T}=0.3$(green), $\tilde{T}=0.4$(red)) with $C=2.0$, $\tilde{Q}=10.0$, $\tilde{\mathcal{A}}=0.005$ when $d=4$.}
    \label{fig:10}
\end{figure}

\begin{figure}
    \centering
    \includegraphics[width=0.85\linewidth]{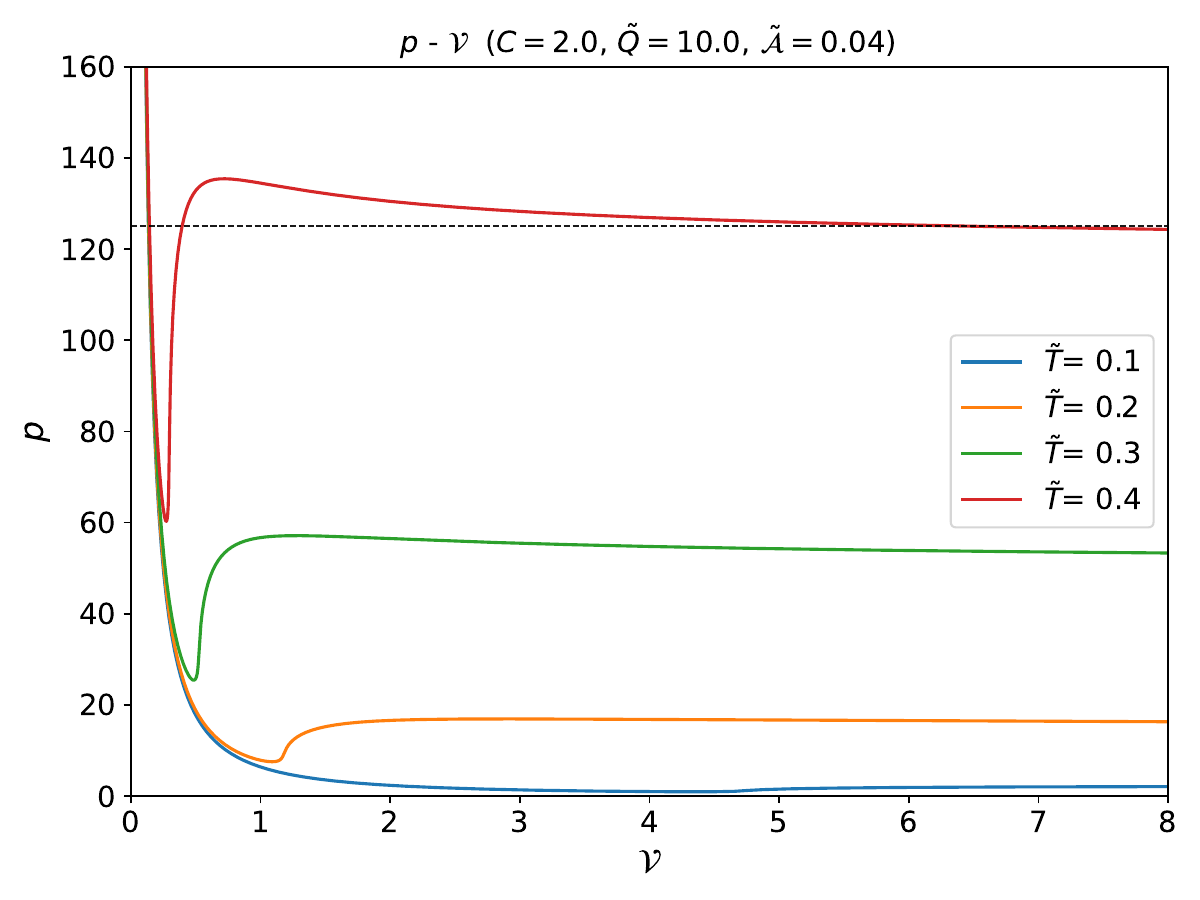}
    \caption{$p-\mathcal{V}$ isotherms curves for different fixed $\tilde{T}$ ($\tilde{T}=0.1$(blue), $\tilde{T}=0.2$(orange),
    $\tilde{T}=0.3$(green), $\tilde{T}=0.4$(red)) with $C=2.0$, $\tilde{Q}=10.0$, $\tilde{\mathcal{A}}=0.04$ when $d=4$. The black dashed line is the equal-area construction line for $\tilde{T}=0.4$.}
    \label{fig:11}
\end{figure}

\begin{figure}
    \centering
    \includegraphics[width=0.85\linewidth]{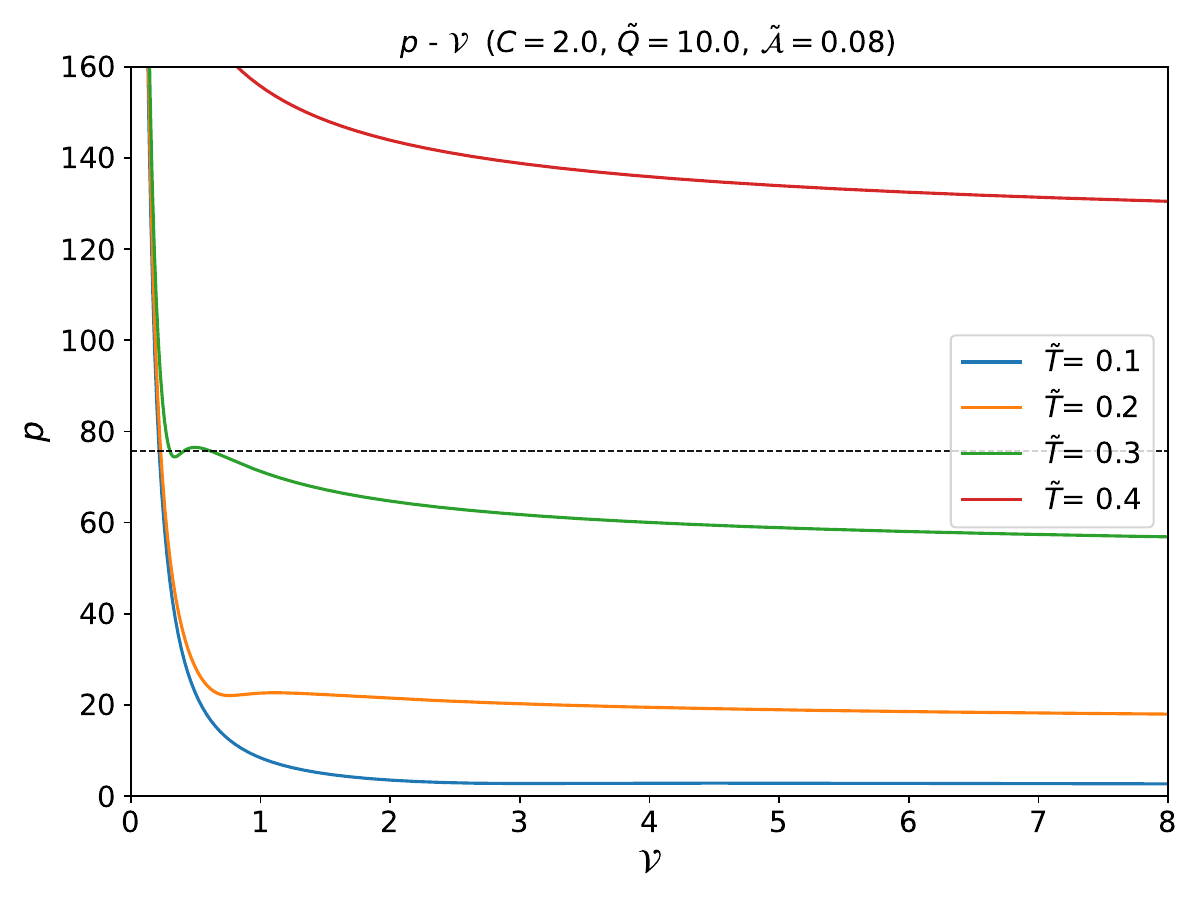}
    \caption{$p-\mathcal{V}$ isotherms curves for different fixed $\tilde{T}$ ($\tilde{T}=0.1$(blue), $\tilde{T}=0.2$(orange),
    $\tilde{T}=0.3$(green), $\tilde{T}=0.4$(red)) with $C=2.0$, $\tilde{Q}=10.0$, $\tilde{\mathcal{A}}=0.08$ when $d=4$. The black dashed line is the equal-area construction line for $\tilde{T}=0.3$.}
    \label{fig:12}
\end{figure}

For $d=4$, the CFT pressure $p$ is a function of $(x,C,\tilde{\mathcal{A}},\mathcal{V})$ as shown in \eqref{eq:2.31}. After setting $y=1$, we can obtain 
\begin{equation}
\begin{split}
p  = \frac{1}{64\pi C x {\mathcal{V}}^{3/2}} \Bigg(
 \tilde{Q}^2 &+ 256\pi^2 C^2 x^4\\&+ 256\pi^2 C^2 x^2 + 256\pi^2 C^2 \tilde{\mathcal{A}}x\Bigg).
\end{split}
\tag{4.2} \label{eq:4.2}
\end{equation}
Numerical results are displayed in figure \ref{fig:10}, \ref{fig:11} and \ref{fig:12}.
The curves exhibit some general features. These figures show that for a fixed temperature, as the value of $\tilde{\mathcal{A}}$ increases, the trough of the $p-\mathcal{V}$ curve gradually rises until the non-monotonic behavior disappears. Moreover, curves at higher temperatures lose this trough feature earlier. For a fixed, moderate value of $\tilde{\mathcal{A}}$ ( figure \ref{fig:11} and \ref{fig:12} ), as the temperature $\tilde{T}$ increases from low to high, the height of the final horizontal plateau of the $p-\mathcal{V}$ curve also becomes progressively higher.

Besides, the isothermal $p-\mathcal{V}$ curves exhibit characteristics similar to those of a van der Waals fluid. In figures \ref{fig:11} and \ref{fig:12}, we have plotted the Maxwell equal-area construction (black dashed line), which indicates the presence of a first-order phase transition during the isothermal process and the existence of a critical temperature.
The specific location of the first-order phase transition point in the phase space can be determined by the following equation:
\begin{equation}
\begin{split}
\int_A^C p d\mathcal{V} = \int_{x_L}^{x_H} p \left( \frac{\partial \mathcal{V}}{\partial x} \right)dx=0.
\end{split}
\tag{4.3} \label{eq:4.3}
\end{equation}
$A$ is the low entropy state and $C$ is the high entropy state. Above this critical temperature, the first-order phase transition ceases to occur. Although the conditions for the critical point of the first-order phase transition, such as the critical temperature, can be obtained by studying the analytical form of the equation \eqref{eq:4.3}, this form is evidently too complex. Therefore, we will not pursue this line of inquiry in this paper.
These results indicate that for a fixed $\tilde{\mathcal{A}}$, there exists a critical $\tilde{T}$, and conversely, for a fixed $\tilde{T}$, there exists a critical $\tilde{\mathcal{A}}$.

\subsection{$\tilde{T}-\tilde{S}$ relationship}

\begin{figure}
    \centering
    \includegraphics[width=0.85\linewidth]{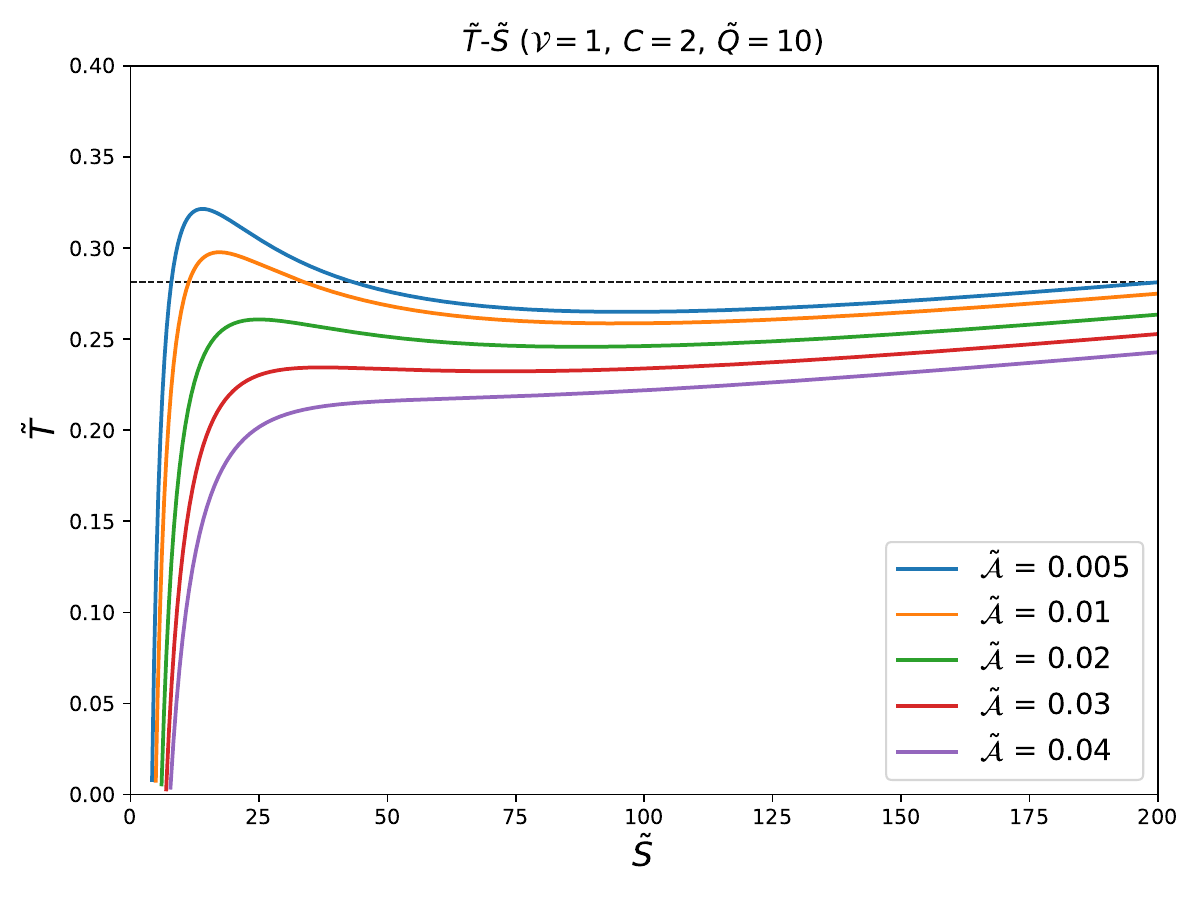}
    \caption{$\tilde{T}-\tilde{S}$ curves for different fixed $\tilde{\mathcal{A}}$ ($\tilde{\mathcal{A}}=0.005$(blue), $\tilde{\mathcal{A}}=0.01$(orange),
    $\tilde{\mathcal{A}}=0.02$(green), $\tilde{\mathcal{A}}=0.03$(red), $\tilde{\mathcal{A}}=0.04$(purple)) with $C=2.0$, ${\mathcal{V}}=1.0$ and $\tilde{Q}=10.0$ when $d=4$. The black dashed line is the equal-area construction line for $\tilde{\mathcal{A}}=0.005$.}
    \label{fig:13}
\end{figure}

For $d=4$, we can plot the $\tilde{T}-\tilde{S}$ curve with fixed $(C, \mathcal{V}, \tilde{Q}, \tilde{\mathcal{A}})$ by using \eqref{eq:2.28} and \eqref{eq:2.29}. The result is shown on figure \ref{fig:13}, with the curve exhibiting characteristics similar to the van der Waals fluid phase transition, which is consistent with the results in ref.~\cite{Yang:2024krx}. 
This is because the curve satisfies Maxwell’s equal-area rule. For small values of $\tilde{\mathcal{A}}$ (blue), an isotherm intersects the curve (black dashed line), enclosing two separate regions of equal area. The temperature corresponding to this isotherm is the temperature at which the first-order phase transition occurs. However, obtaining an analytical expression for this transition temperature is quite complex.  
The first-order phase transition point is given by following equation in the the ensemble at fixed $(C, \mathcal{V}, \tilde{Q}, \tilde{\mathcal{A}})$:
\begin{equation}
\begin{split}
\int_A^C \tilde{T} dS = \int_{x_L}^{x_H} \tilde{T} \left( \frac{\partial S}{\partial x} \right)dx=0.
\end{split}
\tag{4.4} \label{eq:4.4}
\end{equation}
$A$ is the low entropy state and $C$ is the high entropy state. From \eqref{eq:2.28} and \eqref{eq:2.29}, we can expect that there is a difficult expression for \eqref{eq:4.4}. So we are not going to offer such a ugly and lengthy equation.

When $\tilde{\mathcal{A}}$ exceeds a certain value (red, purple), the first-order phase transition reflected by the $\tilde{T}-\tilde{S}$ curve disappears. Anyway, this still does not adequately explain the zeroth-order phase transition discovered from the free energy curve \ref{fig:1}.

\subsection{$\tilde{\Phi}-\tilde{Q}$ relationship}

\begin{figure}
    \centering
    \includegraphics[width=0.85\linewidth]{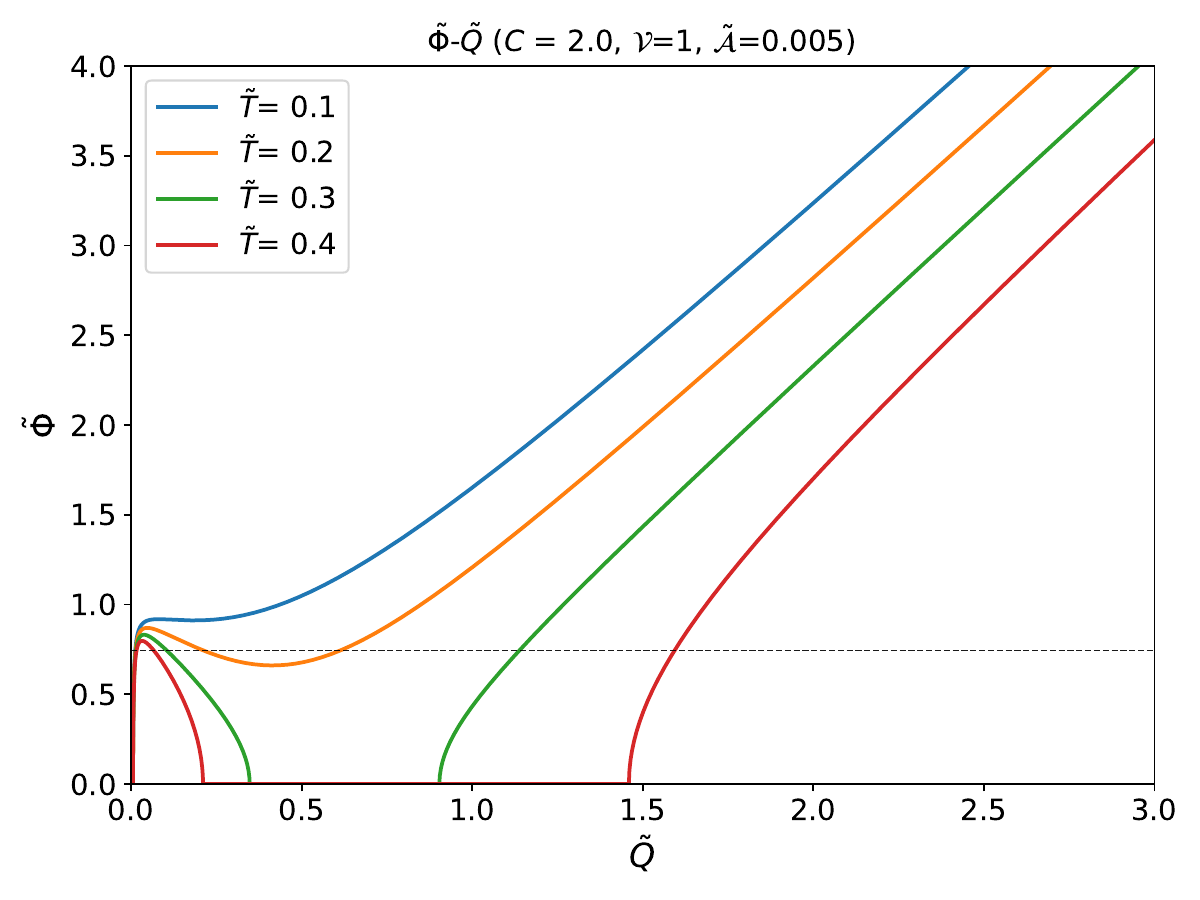}
    \caption{$\tilde{\Phi}-\tilde{Q}$ isotherms curves for different fixed $\tilde{T}$ ($\tilde{T}=0.1$(blue), $\tilde{T}=0.2$(orange),
    $\tilde{T}=0.3$(green), $\tilde{T}=0.4$(red)) with $\mathcal{V}=1.0$, $\tilde{Q}=10.0$, $\tilde{\mathcal{A}}=0.005$ when $d=4$. The black dashed line is the equal-area construction line for $\tilde{T}=0.2$ when $d=4$.}
    \label{fig:14}
\end{figure}

\begin{figure}
    \centering
    \includegraphics[width=0.85\linewidth]{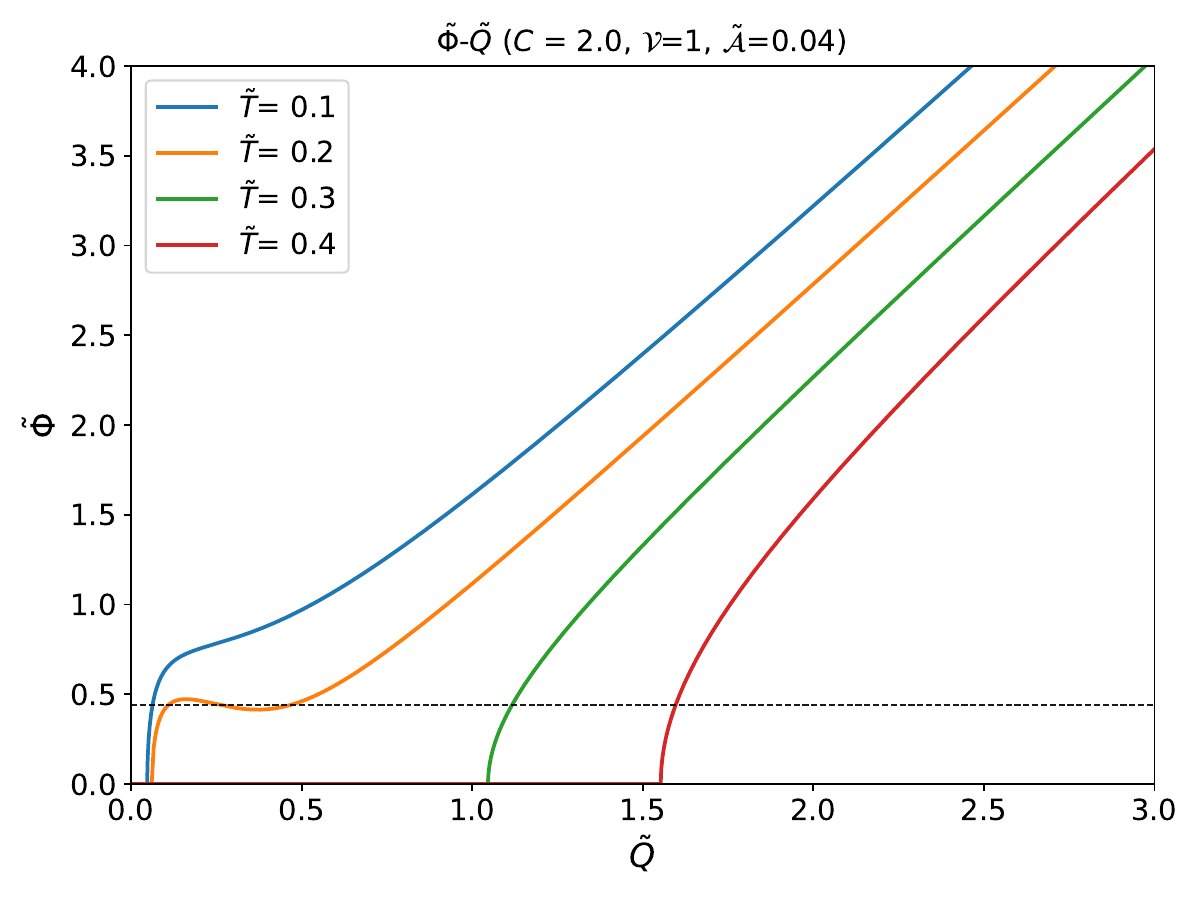}
    \caption{$\tilde{\Phi}-\tilde{Q}$ isotherms curves for different fixed $\tilde{T}$ ($\tilde{T}=0.1$(blue), $\tilde{T}=0.2$(orange),
    $\tilde{T}=0.3$(green), $\tilde{T}=0.4$(red)) with 
    $C=2.0$, $\mathcal{V}=1.0$, $\tilde{\mathcal{A}}=0.04$ when $d=4$. The black dashed line is the equal-area construction line for $\tilde{T}=0.2$ when $d=4$.}
    \label{fig:15}
\end{figure}

\begin{figure}
    \centering
    \includegraphics[width=0.85\linewidth]{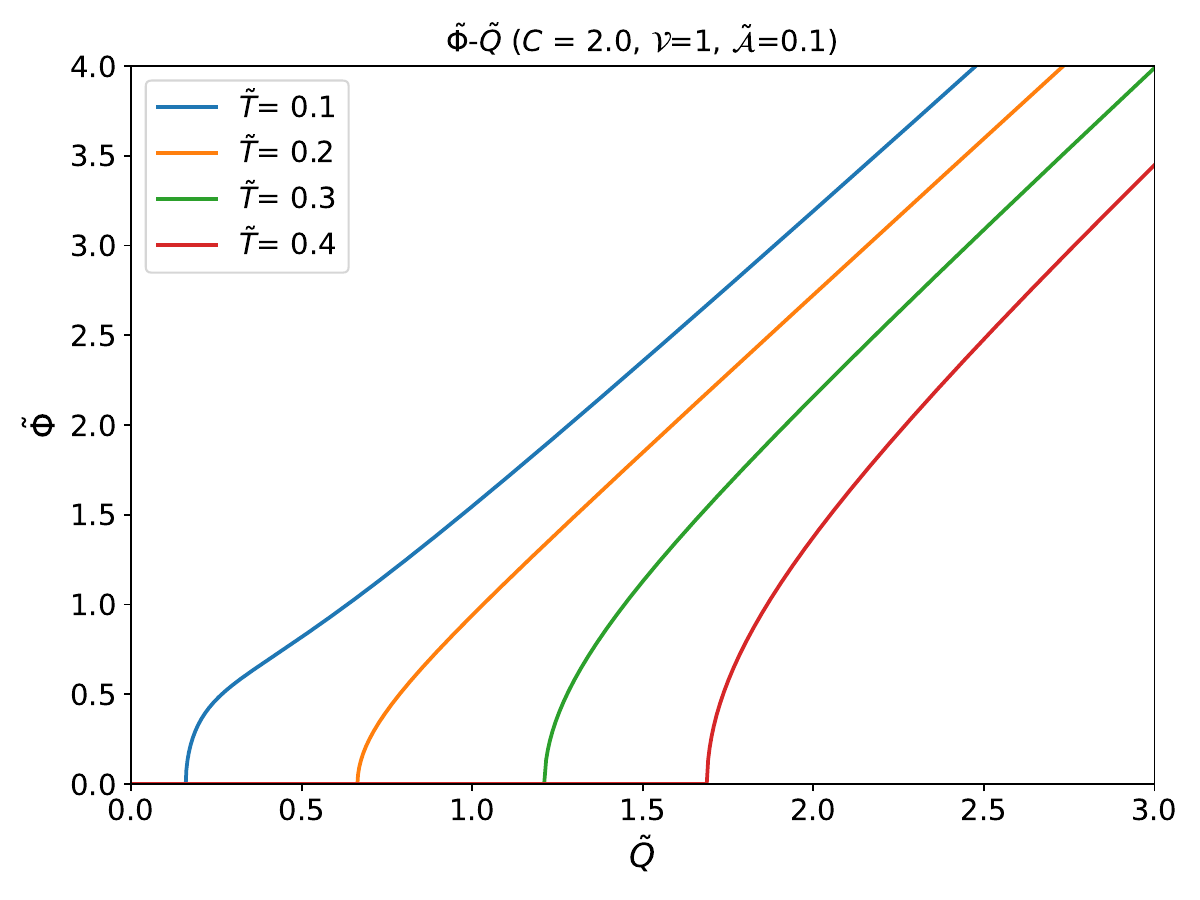}
    \caption{$\tilde{\Phi}-\tilde{Q}$ isotherms curves for different fixed $\tilde{T}$ ($\tilde{T}=0.1$(blue), $\tilde{T}=0.2$(orange),
    $\tilde{T}=0.3$(green), $\tilde{T}=0.4$(red)) with $C=2.0$, $\mathcal{V}=1.0$,  $\tilde{\mathcal{A}}=0.1$ when $d=4$. }
    \label{fig:16}
\end{figure}

For $d=4$, the potential $\tilde{\Phi}$ is a function of $(x,C,\tilde{Q},\mathcal{V})$ as shown in \eqref{eq:2.30}. In our study, we still fix $\mathcal{V}=1$ and $\tilde{Q}=10$, but we varies the value of $\tilde{\mathcal{A}}$ and $\tilde{T}$, which means to investigating the effect of $\tilde{\mathcal{A}}$ on the family of $\tilde{\Phi}-\tilde{Q}$ isotherms. Specifically, for each fixed $\tilde{T}$, we can assign different values to $\tilde{\mathcal{A}}$, solve for $x(\tilde{Q})$ using equation \eqref{eq:3.6}, and then substitute the values of $\tilde{\mathcal{A}}$ and $x$ into equation \eqref{eq:2.30} to obtain different $\tilde{\Phi}-\tilde{Q}$ curves.
Results are displayed in figure \ref{fig:14}, \ref{fig:15} and \ref{fig:16}. 
Here, only the region where $\tilde{\Phi}>0$ and physical $x>0$ is retained. As can be seen from the three figures, for moderate temperatures and values of $\tilde{\mathcal{A}}$ below a certain critical value, a first-order phase transition, determined by Maxwell’s equal-area law, exists during the isothermal process. What's more, for small values of $\tilde{\mathcal{A}}$ (figure \ref{fig:14}), the curve of the high-temperature isothermal process is split into two branches (green curve and red curve). However, for large values of $\tilde{\mathcal{A}}$, all $\tilde{\Phi}-\tilde{Q}$ isotherms start at a certain $\tilde{Q}$ (figure \ref{fig:15} \ref{fig:16}) and lacks the characteristics of a van der Waals fluid phase transition (figure \ref{fig:16}).

\section{Summary and conclusions}
\label{sec:5}

This paper begins in section \ref{sec:2} by presenting the holographic thermodynamic quantities dual to charged AdS black holes in $d=4$ and $d=5$ Gauss-Bonnet gravity. 

In Section \ref{sec:3}, we meticulously plot the CFT free energy curves within the fixed-$\tilde{\mathcal{A}}$ ensemble, from which we identify several critical values of $\tilde{\mathcal{A}}$. The introduction of $\tilde{\mathcal{A}}$ necessitated the artificial constraint $y=1$ to eliminate a redundant intermediate variable. In the $d=4$ case, we examined the influence of both $\tilde{\mathcal{A}}$ and the charge $\tilde{Q}$ on the holographic CFT phase transitions and critical behavior, uncovering several critical points. We then solved the conventional first-order transition criticality equation for both $d=4$ and $d=5$, finding the numerical solution within the $\tilde{\mathcal{A}}$ parameter space under consideration which can explain the critical point. For the case of $d=4$, we determined the correct condition for the critical point of the first-order phase transition by studying the curve of $G$ as a function of $x$. It is noteworthy that when the dependence of the free energy on $\tilde{Q}$ is considered, the type of phase transition gradually transitions from first-order to zeroth-order, which reflects the complexity of the phase structure. Therefore, we were unable to obtain suitable conditions for the critical $\tilde{Q}$ values. However, we do not delve into this point in this paper. All these issues mirror a problem in ref. \cite{Sadeghi:2024ish}, where an artificial constraint was also imposed, but their ensemble’s free energy remained unaffected by $\tilde{\mathcal{A}}\tilde{\alpha}$. Meanwhile, ref.~\cite{Yang:2024krx} investigated critical phenomena for neutral $d=5$ Gauss-Bonnet AdS black holes ensemble at fixed $(C, \mathcal{V}, \tilde{\alpha})$. We discovered that in the $d=5$  ensemble at fixed $(C, \mathcal{V}, \tilde{Q}, \tilde{\mathcal{A}})$, $\tilde{\mathcal{A}}\tilde{\alpha}$ appears only as a constant term, failing to produce the peculiar critical behavior observed in $d=4$. 

What's more, in section \ref{sec:4}, we conduct a numerical study of the $\mu-C$, $p-\mathcal{V}$, $\tilde{T}-\tilde{S}$ and $\tilde{\Phi}-\tilde{Q}$ thermodynamic relations. While $p-\mathcal{V}$, $\tilde{T}-\tilde{S}$ and $\tilde{\Phi}-\tilde{Q}$ curves indicate first-order phase transitions are analogous to the van der Waals fluid phase transition, the $\mu-C$ curves show that $\mu$ has a restricted interval of values. 

This paper merely serves as a starting point. As discussed at the end of section \ref{sec:2}, we employ an unconventional constraint to eliminate the redundant intermediate variable, ensuring the validity of our discussion within this ensemble. This naturally raises a number of questions. However, we are tolerant of the issues that arise, unless there is a specific and explicit prohibition against our choice. The content of this paper is likely to generate significant controversy. Nevertheless, we hope it will prompt further proposals aimed at interpreting or overturning these numerical results.

In our future work, we intend to explore a more suitable interpretation for the dual CFT within this ensemble at fixed $(C, \mathcal{V}, \tilde{Q}, \tilde{\mathcal{A}})$, moving beyond the strategy used in this paper. Our current approach leads to uncommon free energy behavior, notably the appearance of a loop structure. Furthermore, we believe a more systematic explanation is possible, given the resemblance between some of our numerical results and those from the geometric deformation method\cite{panigrahi2025}.
Additionally, analytically solving for the $\tilde{\mathcal{A}}$ and $\tilde{T}$ at the critical point of the van der Waals-like first-order phase transition presented in section \ref{sec:4} is also a highly challenging task.

\begin{acknowledgements}
We are grateful to Junzhong Yang, Ang Gao (School of Physical Science and Technology , BUPT), Chaoqiang Geng (HIAS , UCAS) and for their support, and to Cong Ma (School of Integrated Circuits, BUPT) for his private funding. We are also thanks Jiping Wu's family for their hospitality. 
\end{acknowledgements}

\bibliographystyle{spphys}       
\bibliography{ex}   

\end{document}